\documentclass[a4paper,11pt]{article}
\usepackage{jheppub} 
\usepackage{xcolor}
\definecolor{ForestGreen}{RGB}{34,139,34}
\usepackage[T1]{fontenc}
\usepackage{esint}

\newcommand{\oN}{{\overline{\nabla}}}

\newcommand\Uone{\overset{\scriptscriptstyle 1}{U}\vphantom{U}}

\newcommand\ellzero{\overset{\scriptscriptstyle 0}{\ell}\vphantom{\ell}}
\newcommand\ellone{\overset{\scriptscriptstyle 1}{\ell}\vphantom{\ell}}

\newcommand\thetazero{\overset{\scriptscriptstyle 0}{\theta}\vphantom{\theta}}
\newcommand\thetaone{\overset{\scriptscriptstyle 1}{\theta}\vphantom{\theta}}

\usepackage{cancel}
\usepackage{mdframed,framed}

\newcommand{\beq}{\begin{eqnarray}}
\newcommand{\eeq}{\end{eqnarray}}
\newcommand{\beqn}{\begin{eqnarray}}
\newcommand{\eeqn}{\end{eqnarray}}
\newcommand{\bee}{\begin{equation} \begin{aligned}}
\newcommand{\eee}{ \end{aligned} \end{equation}}
\newcommand{\pa}{\partial}

\newcommand{\cL}{{\cal L}}
\newcommand{\cO}{{\cal O}}
\newcommand{\fL}{\mathfrak{L}}

\usepackage{tikz-cd}
\usepackage{pict2e}
\makeatletter
\newcommand{\variable@rule}[1]{%
  \fontdimen8  
  \ifx#1\displaystyle\textfont3\else
    \ifx#1\textstyle\textfont3\else
      \ifx#1\scriptstyle\scriptfont3\else
        \scriptscriptfont3\relax
  \fi\fi\fi
}
\usepackage{mathtools}

\usepackage{tocloft}
\newcommand{\ve}{\varepsilon}

\newcommand{\cN}{{\cal{N}}}
\newcommand{\cM}{{\cal{M}}}
\newcommand{\cC}{{\cal{C}}}

\newcommand{\bq}{{\overline{q}}}

\newcommand{\bz}{{\overline{z}}}
\newcommand{\oD}{{\overline{D}}}

\newcommand{\rd}{\text{d}}

\newcommand{\chkM}{{\color{red} \,\checkmark\kern-5pt{}_{M}}}

\newcommand{\be}{\begin{equation}}
\newcommand{\ee}{\end{equation}}
\newcommand{\bea}{\begin{eqnarray}}
\newcommand{\eea}{\end{eqnarray}}

\def\pa{\partial}

\usepackage{ragged2e}
\usepackage{blindtext}

\newcommand{\beqa}{\begin{eqnarray}}
\newcommand{\eeqa}{\end{eqnarray}}

\newcommand{\unk}{{\underline{k}}}

\usepackage{tikz-cd}
\usepackage{pict2e}
\makeatletter
\usepackage{mathtools}
\newcommand{\qq}{\qquad}
\newcommand{\unn}{{\underline{n}}}
\newcommand{\uk}{{\underline{k}}}
\def\pa{\partial}
\newcommand{\eqN}{ \ {\stackrel{{\cN}}{=}} \ }

\title{Asymptotic Limit of Null Hypersurfaces}

\author{Luca Ciambelli}
\affiliation{Perimeter Institute for Theoretical Physics,\\
			 31 Caroline St. N., Waterloo ON, Canada, N2L 2Y5}

\emailAdd{ciambelli.luca@gmail.com}

\abstract{We study null hypersurfaces approaching null infinity in asymptotically flat spacetimes within the Bondi-Sachs gauge. The null Raychaudhuri constraint is shown to asymptote to the Bondi mass-loss formula, interpreted as a stress tensor conservation law. This stress tensor, the null Brown-York tensor, yields a Carrollian stress tensor at null infinity from the bulk. Furthermore, we establish that the canonical phase space on finite-distance null hypersurfaces asymptotes to the Ashtekar-Streubel phase space. This connection between finite-distance null physics and null infinity unveils promising insights.}

\begin{document}
\maketitle
\flushbottom

\newpage 

\section{Introduction}

The geometric description of a null (aka lightlike) hypersurface is mathematically intricate, yet the underlying physics is strikingly simple: it is ultra-local and inherently conformal.

Over the past decade, significant efforts have focused on unraveling the physics of null hypersurfaces at finite distance in the bulk. Building on the foundational works of \cite{Henneaux1979a} and the Rigging construction \cite{Mars:1993mj} (see also \cite{Gourgoulhon:2005ng}), the study of null physics has advanced through the framework of Carrollian geometry \cite{Hartong:2015xda, Ciambelli:2018wre, Ciambelli:2019lap, Redondo-Yuste:2022czg, Freidel:2022bai, Ciambelli:2023mir}. This has enabled a deeper understanding of the mathematical complexities associated with degenerate metrics and non-Levi-Civita connections. Furthermore, in a series of works \cite{Chandrasekaran:2018aop, Chandrasekaran:2020wwn, Chandrasekaran:2021hxc, Chandrasekaran:2021vyu}, the null analogue of the Brown-York stress tensor \cite{brown1993quasilocal} was constructed, filling a critical gap in the literature.

In parallel, progress has been made in understanding null infinity from a Carrollian viewpoint \cite{Ciambelli:2018wre, Campoleoni:2018ltl, Mason:2023mti, Alday:2024yyj}, and its connection to the celestial holography program \cite{Donnay:2022aba, Donnay:2022wvx}.\footnote{For a review of the celestial holography program, see \cite{Strominger:2017zoo, Raclariu:2021zjz, Pasterski:2021rjz, Pasterski:2023ikd, Donnay:2023mrd}, and references therein.} Despite these advances, two key elements remain missing in flat-space holography: the geometric description of the bulk-to-boundary limit by foliating spacetime with hypersurfaces akin to the boundary one, and, related to the previous point, the construction of a null stress tensor derived from the bulk action in the asymptotic limit.

In AdS/CFT, the foliation of AdS with timelike hypersurfaces (${\cal B}$ in the figure below) is central to the construction of the Balasubramanian-Kraus stress tensor \cite{brown1993quasilocal, Balasubramanian:1999re, Emparan:1999pm, deHaro:2000vlm}, providing insights into the renormalization of this stress tensor and the construction of boundary responses to a source. Moreover, this foliation reveals how the holographic coordinate can be understood as an RG flow \cite{deBoer:1999tgo, Bianchi:2001kw, Skenderis:2002wp}. A natural question then arises: how does flat-space holography behave when viewed through the lens of hypersurfaces (${\cal N}$ in the figure 1 below) that share the geometric structure of future null infinity, as illustrated below?
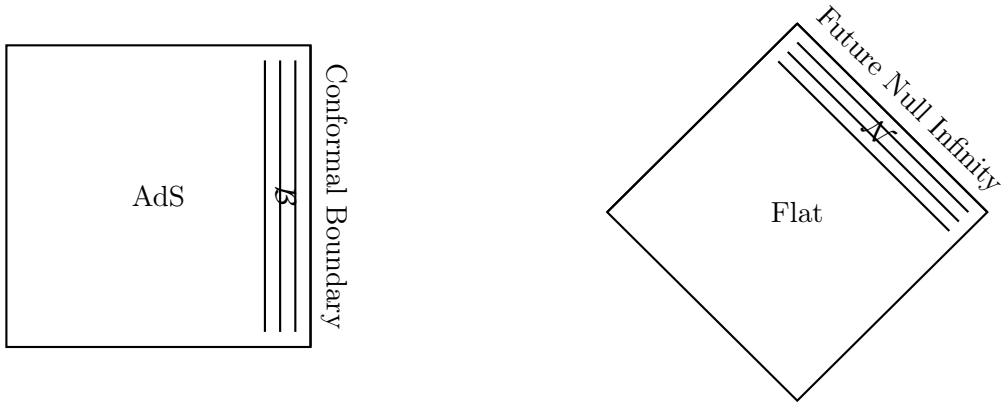
\begin{figure}[htbp]
    \centering
    \begin{minipage}{0.45\textwidth}
        \centering
        \begin{tikzpicture}[scale=2]
\draw[thick] (0,0) rectangle (2,2);
\draw[thick] (2,2) -- (2,0) node[midway, above, sloped] {{\text{Conformal Boundary}}};

\draw[thick] (1.9,0.1) -- (1.9,1.9) node[midway, above, sloped] {};
\draw[thick] (1.8,0.1) -- (1.8,1.9) node[midway, above, sloped] {};
\draw[thick] (1.7,1.9) -- (1.7,0.1) node[midway, above, sloped] {{$\cal B$}};
\node at (1,1) {AdS};
\end{tikzpicture}
    \end{minipage}
    \hfill
    \begin{minipage}{0.45\textwidth}
        \centering
        \begin{tikzpicture}[scale=2.5]

\draw[thick] (-1,-1) -- (0,0) -- (1,-1) -- (0,-2) -- cycle;

\draw[thick] (0,0) -- (1,-1) node[midway, above, sloped] {{\text{Future Null Infinity}}};

\draw[thick] (-0.05,-0.15) -- (0.85,-1.05) node[midway, above, sloped] {};
\draw[thick] (-0.0,-0.1) -- (0.9,-1.0) node[midway, above, sloped]{};
\draw[thick] (-0.1,-0.2) -- (0.8,-1.1) node[midway, above, sloped]{{$\cal N$}};
\node at (0,-1) {Flat};

\draw[thick] (0,0) -- (-1,-1) node[midway, above, sloped]{};
\end{tikzpicture}
    \end{minipage}
    \vspace{0.5cm}
    \caption{Foliations of asymptotically AdS (left) and asymptotically flat (right) spacetimes with family of hypersurfaces with the same causal nature as the asymptotic boundary of interest. For AdS, the boundary is timelike, while for flat, the boundary is null.}
\end{figure}

Foliating the bulk with null hypersurfaces in flat space not only allows us to establish a connection between finite-distance Carrollian physics and the asymptotic behavior at null infinity, but also lays the groundwork for a unified framework. Indeed, a universal understanding of the thermodynamic and hydrodynamic properties of Carrollian fluids dual to gravity is still lacking, although they are crucial in understanding both black hole horizons \cite{Donnay:2019jiz} and null infinity \cite{Ciambelli:2018wre}. Specifically, while on a finite-distance null hypersurface Einstein equations (and in particular the Raychaudhuri constraint) have been understood as the conservation law of the null Brown-York stress tensor \cite{Chandrasekaran:2021hxc, Chandrasekaran:2021vyu}, such an understanding is currently missing for the asymptotic Einstein equations, and in particular for the Bondi mass-loss formula. By demonstrating that the Raychaudhuri equation on our family of null hypersurfaces exactly asymptotes to the Bondi mass-loss formula, we achieve in this manuscript a deeper understanding of the latter. This can be seen as formulating a membrane paradigm for asymptotic null infinity, in the spirit of \cite{Damour1979, Price:1986yy}.

Recent work has begun exploring the construction of a flat-space stress tensor analogous to the AdS/CFT one \cite{Riello:2024uvs, Bhambure:2024ftz}. However, these studies consider a timelike hypersurface in the bulk, the stretched horizon, that asymptotes to null infinity. The key novelty of the present work lies in the direct consideration of a family of null hypersurfaces within the bulk, providing a more direct and profound connection between bulk physics on null hypersurfaces and the null conformal boundary. In a similar vein, our procedure allows us to relate the finite-distance gravitational phase space and dynamics discussed in \cite{Reisenberger:2007ku, Wieland:2017zkf, Chandrasekaran:2018aop, Adami:2020ugu, Adami:2021kvx, Odak:2023pga, Chandrasekaran:2023vzb, Ciambelli:2023mir} to null infinity. Indeed, in this work we will demonstrate that this phase space smoothly and precisely asymptotes to the Ashtekar-Streubel phase space \cite{Ashtekar1981}, opening the door to a wide range of future investigations, which we summarize in the Conclusions. We regard the present work as an initial exploration, setting the stage and vocabulary for future works.

Here is the road-map of the paper. In Section \ref{sec:bondi} we present the Bondi-Sachs gauge and the asymptotic Einstein equations of motion. We then study null hypersurfaces in this gauge in Section \ref{sec:asymptotic0}. We begin in \ref{sec:constructing} by constructing a null hypersurface in the bulk and reviewing how to induce from the bulk the geometric data via the Rigging projector. We then send this hypersurface toward future null infinity in subsection \ref{sec:asymptotic}. This allows us to read off the intrinsic Carrollian data order by order toward the boundary, which we do in \ref{sec:intrinsic}. To ensure a smooth evolution of the section, we defer details to Appendix \ref{AppA}. Eventually, we compute the leading order Einstein equations as intrinsic constraints in subsection \ref{sec:einstein}, demonstrating how they involve the boundary metric only. In Section \ref{sec:simplified}, we essentially reproduce the same analysis of the previous section, in the restricted framework where the boundary metric is time independent. We first discuss the intrinsic data in \ref{sec:intrinsic2} and then discuss the subleading Einstein equations in subsection \ref{sec:einstein2}. Here, we prove that the sub-sub-leading order of the Raychaudhuri constraint is precisely the Bondi mass-loss formula. We then construct the holographic stress tensor (subsection \ref{sec:holographic}), and study it in a further simplified framework. This allows us to understand the Bondi mass as the energy density of the asymptotic Carrollian fluid, while the News tensor acts as a viscous tensor. Eventually, we match the finite-distance phase space and the Ashtekar-Streubel phase space in subsection \ref{sec:covariant}. As emphasized, this paper represents a crossroads for further exploration. We summarize these directions, along with a recap of the main results, in Section \ref{sec:final}.

\section{Bondi-Sachs Gauge}\label{sec:bondi}

The Bondi-Sachs (BS) line element is suitable to describe the asymptotic structure of an asymptotically flat spacetime \cite{Bondi, Sachs:1961zz}, see also \cite{Barnich:2010eb, Compere:2018ylh, Campiglia:2020qvc, Freidel:2021fxf}. Working in $4$ spacetime dimensions, and considering coordinates $y^\mu=(r,x^a)=(r,u,\sigma^A)=(r,u,z,\bz)$, this line element is given by
\beq\label{BS}
\rd s^2=-2e^{2\beta}\rd u(\rd r+F \rd u) +g_{AB}(\rd \sigma^A-U^A \rd u)(\rd \sigma^B-U^B \rd u).
\eeq
The conformal boundary is located at $r\to \infty$, where the metric of the manifold $\cM$ displays a pole of order $2$ in its spatial components. To accommodate this, we consider here the following falloffs
\beqn\label{fo1}
F&=&\bar F-\frac{m}{r}+\cO(r^{-2})\\
\beta&=& \frac{\bar\beta}{r^2}+\cO(r^{-3})\\
g_{AB}&=&r^2\bar q_{AB}+rC_{AB}+\frac1{4}\bar q_{AB} C_{CD}C^{CD}+\frac1{r}E_{AB}+\cO(r^{-2})\\
U^A&=&\frac{\bar U^A}{r^2}+\frac{U_1^A}{r^3}+\cO(r^{-4}),\label{fo2}
\eeqn
where $U_1^A=-\frac2{3}\bar q^{AB}(\bar P_B+C_{BC}\bar U^C+\pa_B \bar\beta)$ and $\bar q^{AB}$ is the inverse of the boundary spatial metric $\bar q_{AB}$, used to raise spatial indices of boundary tensors. The quantity $m$ is the Bondi mass, $C_{AB}$ is the asymptotic shear, and $\bar P_A$ is the generalization of the angular momentum.
Solving Einstein equations, the quantities appearing in the expansion are related via
\beq\label{eom}
\bar\beta+\frac1{32}C_{AB}C^{AB}=0\qquad \bar R=4\bar F\qquad \bar U^A+\frac12  \oN_B C^{AB}=0\,,
\eeq
where $\bar R$ is the Ricci scalar of the spatial boundary metric $\bar q_{AB}$, while $\bar \oN_A$ is its Levi-Civita covariant derivative, such that $\oN_A \bar q_{BC}=0$. We emphasize that in \eqref{fo1} we are making a stringent assumption on $F$. Indeed, if we authorize an order $r$ term, $F=r K+\bar F-\frac{m}{r}+\cO(r^{-2})$, then this term allows us to have an undetermined time dependent conformal factor in the spatial boundary metric, see \cite{Barnich:2010eb}. However, this extra term drastically complicates the description of a family of null hypersurfaces. Since we are ultimately interested in a time-independent boundary metric, we set $K=0$ from the beginning and leave this generalization for the future.

There are furthermore the constraints along the null generator. Given our expansion \eqref{fo1}, the leading order one is a second-derivative condition solved setting the boundary metric to be time independent
\beq\label{dq}
\pa_u \bar q_{AB}=0.
\eeq
The subleading orders are the Bondi mass-loss formula
\beq\label{bmlf}
\pa_u m=\frac14 \oN_A \oN_B N^{AB}+\frac18 \bar \Delta \bar R-\frac18 N_{AB}N^{AB},
\eeq
and the "angular-momentum" equation
\beq\label{pauP}
\pa_u \bar P_A&=&\oN_A m+\frac18 \oN_A (C^{BC}N_{BC})+C_{AB}\pa^B \bar F +\frac{1}{4} \oN_C\left(\oN_A \oN_B C^{B C}-\oN^C \oN^B C_{A B}\right)\nonumber\\
&&+\frac{1}{4} \oN_B\left(N^{B C} C_{A C}-C^{B C} N_{A C}\right)-\frac{1}{4} N^{B C} \oN_A C_{B C}.
\eeq
In the following, we will not impose these three constraints: our goal is to derive them as the asymptotic limit of the Einstein equations projected onto an asymptotic null hypersurface.

For convenience, we report the non-vanishing components of the metric and its inverse
\beq
&g_{ur}=-e^{2\beta},\qq g_{uu}=g_{AB}U^AU^B-2e^{2\beta}F,\qq g_{AB},\qq g_{Au}=-g_{AB}U^B&\\
&g^{ur}=-e^{-2\beta},\qq g^{rr}=2e^{-2\beta}F,\qq g^{AB}, \qq g^{Ar}=-e^{-2\beta}U^A,&
\eeq
where $g^{AB}g_{BC}=\delta^A_C$.

This is our starting point. In the following, we will describe a family of null hypersurfaces in the gauge \eqref{BS}, and its asymptotic limit.

\section{Asymptotic Null Hypersurfaces}\label{sec:asymptotic0}

In this section, we first define a null hypersurface, and study how to induce from the bulk its geometric properties, using the null Rigging construction of \cite{Mars:1993mj}. We then send this hypersurface toward the conformal boundary, expanding the various quantities in the BS gauge. We recast these bulk tensors intrinsically on the hypersurface, and use the Carrollian language that we developed in \cite{Ciambelli:2023mir, Ciambelli:2023mvj}\footnote{This is a culmination of previous works on finite-distance Carrollian physics and geometry, \cite{Henneaux1979a, Duval:2014uva, Hartong:2015xda, Hopfmuller:2016scf, Hopfmuller:2018fni, Chandrasekaran:2018aop, Donnay:2019jiz, Ciambelli:2019lap}.} to understand the expansion and shear of the null generators. This then allows us to study the intrinsic Einstein constraints on the hypersurface, i.e., the Raychaudhuri and Damour equations. We do so using the recently introduced null Brown-York stress tensor \cite{Chandrasekaran:2020wwn, Chandrasekaran:2021hxc, Chandrasekaran:2021vyu, Freidel:2022bai, Ciambelli:2023mir}, suitable to formulate these equations of motion as intrinsic conservation laws. 

\subsection{Constructing Null Hypersurfaces}\label{sec:constructing}

A null hypersurface $\cN$ inside $\cM$ is defined specifying its normal $1$-form
\beq \label{norn}
n=-\rd(r-r_{\cN}(u,\sigma)),
\eeq
such that its bulk location is $r=r_{\cN}(u,\sigma)$. Note that we must allow for a non-trivial function $r_{\cN}(u,\sigma)$, otherwise the hypersurface at $r=cnst$ is timelike.\footnote{Actually, $r=cnst$ is timelike only asymptotically when the boundary spatial topology is a $2$-sphere.} Therefore, the embedding map from $\cN$ to $\cM$ is
\beq
\phi:{\cN}\to {\cM},\qq \phi: (u,\sigma)\mapsto (r_{\cN}(u,\sigma),u,\sigma).
\eeq

The metric-dual vector field of the normal $1$-form in BS gauge is
\beq\label{vectn}
\unn=-e^{-2\beta}(2F+(\pa_u +U^A\pa_A) r_{\cN})\pa_r+e^{-2\beta}\pa_u+(g^{AB}\pa_B r_{\cN}+e^{-2\beta}U^A)\pa_A.
\eeq
Its norm is therefore
\beq\label{norm}
|n|^2=n^\mu n_\mu=2e^{-2\beta}(F+(\pa_u+U^A\pa_A)r_{\cN})+\pa_Ar_{\cN}g^{AB}\pa_Br_{\cN}.
\eeq
To ensure that the hypersurface is null, we must solve $|n|^2=0$. We will do so perturbatively toward the conformal boundary. 

Prior to that, we can set up the null Rigging construction \cite{Mars:1993mj}, see also \cite{Gourgoulhon:2005ng}. Consider the auxiliary vector field
\beq
\unk=-\pa_r.
\eeq
One readily obtains
\beq\label{k}
k=e^{2\beta}\rd u,\qq k^\mu k_\mu=0,\qq k^\mu n_\mu=1.
\eeq

The Rigging projector
\beq\label{Riggpro}
\Pi_\mu{}^\nu=\delta_\mu^\nu-n_\mu k^\nu,
\eeq
projects bulk tensors to the hypersurface. It satisfies by construction
\beq
\Pi_\mu{}^\nu \Pi_\nu{}^\rho=\Pi_\mu{}^\rho
\eeq
and
\beq
\Pi_\mu{}^\nu k_\nu=k_\mu\quad \Pi_\mu{}^\nu n_\nu=0\quad k^\mu\Pi_\mu{}^\nu=0 \quad n^\mu\Pi_\mu{}^\nu=n^\nu-|n|^2 k^\mu.
\eeq
In the last expression, we kept explicit the norm of $n$, as we will set that to zero only perturbatively near the boundary. 

Clearly, selecting a different vector $\uk$ would have induced a different one-form $k$ on the surface. This is the sense in which $k$ is a connection, not uniquely determined intrinsically. In the intrinsic Carrollian language, $k$ is the Carrollian Ehresmann connection \cite{Ciambelli:2019lap}, while the vector $n^\mu$ becomes the Carrollian vector field $\ell^a$ on the hypersurface \cite{Henneaux1979a}. 

Using $\eqN$ to indicate equalities holding on $\cN$, the intrinsic Carrollian geometric data are the Carrollian vector field 
\beq\label{lC}
\ell^a\eqN n^\mu \Pi_\mu{}^a,
\eeq
the induced degenerate metric 
\beq\label{qp}
q_{ab}\eqN \Pi_a{}^\mu \Pi_b{}^\nu g_{\mu\nu},
\eeq
and the Ehresmann connection
\beq\label{kC}
k_a \eqN \Pi_a{}^\mu k_\mu.
\eeq

We furthermore introduce two important quantities, the inaffinity $\kappa$ and the H\'aji\u{c}ek connection $\pi_a$, defined as
\beq\label{defkappa}
\kappa\eqN k_\nu n^\mu \nabla_\mu n^\nu, \qquad \pi_a \eqN q_a{}^\nu k_\rho \nabla_\nu n^\rho\,,
\eeq
where $\nabla_\mu$ is the bulk Levi-Civita connection, and we recall that $x^a=(u,\sigma^A)$ are the intrinsic coordinates on $\cN$. These quantities are the fundamental ingredients of the intrinsic Carrollian geometry, reviewed in Appendix \ref{AppA}. 

We conclude this subsection noticing that the BS gauge is designed to study timelike hypersurfaces near the asymptotic boundary ($r=cnst$), rather than null. Of course, null hypersurfaces can be studied in this gauge, but their treatment becomes technically challenging. We decided nonetheless to use the BS gauge as this will make use of a language and a set of tools that are familiar in the community.

\subsection{Asymptotic Expansion}\label{sec:asymptotic}

We wish to impose the vanishing of \eqref{norm} perturbatively. First of all, we require that $\cN$ asymptotes to future null infinity, which imposes\footnote{Had we not assumed $K=0$ (see discussion below \eqref{eom}), we would have had to multiply the leading $\lambda$ order by an arbitrary function on the boundary, complicating the analysis that follows.}
\beq\label{pos}
r_{\cN}(u,\sigma)=\lambda+c_0(u,\sigma)+\frac{c_1(u,\sigma)}{\lambda}+\frac{c_2(u,\sigma)}{\lambda^2}+\cO(\lambda^{-3}),
\eeq
such that the asymptotic limit is reached sending $\lambda$ to infinity.\footnote{Note that, by construction, this means that the various functions $c_0, c_1, \dots$ appearing in the expansion cannot explicitly depend on $\lambda$.} Then, using (\ref{fo1}-\ref{fo2}) and asymptotically setting \eqref{norm} to zero gives 
\beq\label{conds}
\pa_u c_0=-\bar F,\qq \pa_u c_1=m,
\eeq
up to order $\lambda^{-2}$ excluded. Therefore, the boundary curvature and the Bondi mass set the position of a null hypersurface near the conformal boundary. As we will see, this comes about because they provide energy to the system, and thus they bend the null rays. Note furthermore that $c_0$ and $c_1$ are specified by the curvature and Bondi mass profiles on the entirety of $\cN$. Importantly, a finite-distance null hypersurface generally develops caustics at finite null time. As it is well-known, this implies that, while there certainly exist inextensible null geodesics, the non-trivial curvature and topology of a spacetime makes it impossible to describe a twist-free (integrable) affine null hypersurface globally. This is readily seen in the conditions we found above, \eqref{conds}, since a simple integral from $-\infty$ to $\infty$ may lead to a divergency. The formation of caustics is a physical and unavoidable feature, see e.g. \cite{Gadioux:2023pmw} and \cite{Ciambelli:2024swv}, stemming directly from the Raychaudhuri equation. In a general spacetime, the only caustic-free null hypersurfaces are black hole horizons (thanks to the teleological boundary condition) and asymptotic null infinity. Therefore, in order to connect our finite-distance analysis to conformal infinity, we must restrict to a finite range of the coordinate $u$. Note that this fact persists in any gauge or coordinate choice, as caustics are physical, and lead to the impossibility of defining a globally-integrable null hypersurface. Our analysis will anyway pertain to either local or quasi-local quantities. Finally, we remark that a more direct way to study this family of null hypersurfaces could have been to change coordinates, such that \eqref{norm} reads $n=-\rd v(r,u,\sigma)$. This is indeed what is happening order by order in the asymptotic expansion. Then, the integration constants we found can be understood as coming from the differential equations to be solved to find the exact diffeomorphism. Let us show this more explicitly, as it will help the reader in familiarizing with the $\lambda$ expansion, and its geometric meaning. Consider equation \eqref{norm}, with the asymptotic expansion \eqref{pos}. Then, on the null hypersurface, one has $r=r_{\cN}=\lambda+c_0+\frac{c_1}{\lambda}+\dots$. If one inverts this asymptotic series, the first orders are $\lambda=r-c_0-\frac{c_1}{r}-\cO(r^{-2})$, where now the expansion is in the radial coordinate $r$. Looking at \eqref{conds}, and momentarily assuming \eqref{eom} with the boundary a sphere, and thus $\bar R=2$, we have that $c_0=-\frac{u}{2}$ plus potentially an integration constant. Therefore, $\lambda=r+\frac{u}{2}-\frac{c_1}{r}-\cO(r^{-2})$. This is exactly the asymptotic expansion in $r$ of the outgoing null coordinate $v$. Therefore, $\lambda$ is the other null coordinate $v$, which is organizing the asymptotic expansion of this family of null hypersurfaces. If one starts from $n=-\rd v$ and expand $v$ in powers of $r$, one finds exactly the asymptotic series written above for $\lambda$. This clarifies the physical and geometric meaning of $\lambda$ in the asymptotic set up of this paper.

The $1$-form \eqref{norn} then acquires the asymptotic expansion
\beq\label{nexp}
n=-(\rd r+(\bar F-\frac{m}{\lambda})\rd u)+(\pa_A c_0 +\frac{\pa_A c_1}{\lambda})\rd \sigma^A+\cO(\lambda^{-2}).
\eeq
The first part of \eqref{nexp} is the asymptotic expansion of $n_1=-(\rd r+F\rd u)$, which is a natural $1$-form stemming from the BS line element \eqref{BS}. One could have taken the $1$-form $n_1$ as the starting point to construct the null hypersurface, which is indeed what is done in \cite{Kapec:2016aqd}.  However, to go deeper into the bulk, the extra spatial derivative terms in \eqref{nexp} are crucial and should be included to ensure Frobenius theorem, and thus integrability. The integrability condition is $n\wedge \rd n=0$, which is by construction automatically satisfied by \eqref{norn}, whereas one has
\beq
n_1\wedge \rd n_1= \pa_A F \ \rd r\wedge \rd \sigma^A\wedge \rd u,
\eeq
which cannot be set to zero in general as it would imply that the boundary metric and the Bondi mass are angle-independent quantities. Relatedly, one can easily show that the metric-dual of $n_1$ is not an inaffine geodesic vector, that is, $n_1^\mu\nabla_\mu n_1^\nu$ is not proportional to $n_1^\nu$. This is the reason why we have set up the geometric construction starting from the $1$-form \eqref{norn}, which is by construction integrable. Note that one can nonetheless chose to work with $n_1$, as long as only the first two terms in the asymptotic expansion are considered. 

It turns out that $n^\mu$ is asymptotically an affine geodesic vector field. Indeed, expanding the inaffinity equation \eqref{defkappa}, we find
\beq\label{kappa}
\kappa=\cO(\lambda^{-2}).
\eeq
This will be important for the asymptotic limit of the Raychaudhuri and Damour equations. 

We can now expand all the quantities introduced in the previous section, that will define the geometric data on the null hypersurface. We keep all the terms up to the first one in which $c_2$ appears, and/or subleading terms of those displayed in (\ref{fo1}-\ref{fo2}) enter. From \eqref{vectn}, we get
\beq
&n^r=-\bar F+\frac{m}{\lambda}+\cO(\lambda^{-2}),\qq n^u=1-\frac{2\bar\beta}{\lambda^2}+\cO(\lambda^{-3}),&\\ &n^A=\frac{1}{\lambda^2}(\bar U^A+\bar q^{AB}\pa_B c_0)+\frac{1}{\lambda^3}(U_1^A-2 c_0 \bar U^A&\nonumber\\
&-C^{AB}\pa_B c_0+\bar q^{AB}(\pa_B c_1-2 c_0 \pa_B c_0))+\cO(\lambda^{-4}).&
\eeq
From \eqref{k}, we readily obtain
\beq
k=(1 + \frac{2\bar\beta}{\lambda^2}+\cO(\lambda^{-3}))\rd u,
\eeq
whereas the projected metric \eqref{qp} gives
\beq
q_{r\mu} & =& 0,\\
q_{uu}  & =& \cO(\lambda^{-2}),\\
q_{u A} & =& -\bar{U}_A-\pa_A c_0 - \frac{1}{\lambda}\left( C_{AB}\bar{U}^B+U_{1A}+\pa_A c_1\right)+\cO(\lambda^{-2}),\\
q_{AB} & =& \lambda^{2} \bar q_{AB} + \lambda (C_{AB}+2 c_0 \bar q_{AB})  + \cO(\lambda^0),
\eeq
where we defined $\bar U_A=\bar q_{AB}\bar U^B$ and $U_{1A}=\bar q_{AB}U_1^B$. 

\subsection{Intrinsic Carrollian Data}\label{sec:intrinsic}

We are ready to induce from the bulk the Carrollian geometric data characterizing this asymptotic null hypersurface.

The Carrollian vector field \eqref{lC} has components
\be\label{lu}
\ell^u=1-\frac{2\bar\beta}{\lambda^2}+\cO(\lambda^{-3}),
\ee
and
\be\label{lA}
\ell^A=\frac{1}{\lambda^2}\ellzero^A+\frac{1}{\lambda^3}\ellone^A+\cO(\lambda^{-4}),
\ee
with
\beq
\ellzero^A&=&\bar U^A+\bar q^{AB}\pa_B c_0\\
\ellone^A&=&U_1^A-2 c_0 \bar U^A-C^{AB}\pa_B c_0+\bar q^{AB}(\pa_B c_1-2 c_0 \pa_B c_0).
\eeq

The Ehresmann connection \eqref{kC} has only the $u$ component, given by
\beq\label{ku}
k_u=1 + \frac{2\bar\beta}{\lambda^2}+\cO(\lambda^{-3}),
\eeq
and one can see by direct inspection that, as required (see the intrinsic description of Carrollian geometry in appendix \ref{AppA}), $k_a\ell^a=1$, up to the desired $\lambda$-order.

The induced degenerate metric \eqref{qp} is $q_{ab}$, with
\beq
q_{uu}  & =& \cO(\lambda^{-2}),\\
q_{u A} & =& -\bar{U}_A-\pa_A c_0 - \frac{1}{\lambda}\left( C_{AB}\bar{U}^B+U_{1A}+\pa_A c_1\right)+\cO(\lambda^{-2}),\\
q_{AB} & =& \lambda^{2} \bar q_{AB} + \lambda (C_{AB}+2 c_0 \bar q_{AB})  + O(\lambda^0).\label{qdd}
\eeq
From this, we can also check intrinsically that $\ell^aq_{ab}=0$.

On the null hypersurface, the projector to the space orthogonal to $k$ and $\ell$ is 
\beq
q_a{}^b=\delta_a^b-k_a\ell^b.
\eeq
Explicitly, we obtain
\beq\label{qdU}   
q_A{}^B = \delta_A^B \qquad
q_a{}^u =0,
\eeq
and $q_u{}^A=-\ell^A$, that is,
\beq\label{quA}
 q_u{}^A&=&-\frac{1}{\lambda^2}(\bar U^A+\bar q^{AB}\pa_B c_0)\\
 &&-\frac{1}{\lambda^3}(U_1^A-2 c_0 \bar U^A-C^{AB}\pa_B c_0+\bar q^{AB}(\pa_B c_1-2 c_0 \pa_B c_0))+\cO(\lambda^{-4}).
\eeq
We can verify explicitly that $q_a{}^b$ satisfies its defining equations
\beq
q_a{}^bk_b=0=\ell^aq_a{}^b,\qquad q_a{}^bq_{bc}=q_{ac}, \qquad q_a{}^bq_b{}^c=q_a{}^c.
\eeq

We now compute the expansion tensor
\beq\label{exo}
\theta_{ab}=\frac12 \cL_{\ell}q_{ab}.
\eeq
Its components are 
\beq
\theta_{AB}&=&\lambda^2\thetazero_{AB}+\lambda\thetaone_{AB}+\cO(1)\\
\theta_{uA}&=&-\ellzero^B\thetazero_{BA}-\frac1{\lambda}\left(\ellone^B\thetazero_{BA}+\ellzero^B\thetaone_{BA}\right)+\cO(\lambda^{-2}) \\
\theta_{uu}&=&\cO(\lambda^{-2}),
\eeq
where we introduced
\beq\label{theta0}
\thetazero_{AB}=\frac12  \pa_u \bq_{AB}\qquad \thetaone_{AB}=c_0 \pa_u \bq_{AB}+\frac12 N_{AB}-\bar F \bq_{AB},
\eeq
with $N_{AB}=\pa_u C_{AB}$ the News tensor.
By design, $\theta_{ab}$ is orthogonal to $\ell^a$, and indeed we verify
\beq
\ell^a\theta_{ab}=\cO(\lambda^{-2}).
\eeq

The crucial ingredients needed to describe a null hypersurface are the shear and expansion of its null generators. These are encoded in the traceless and trace parts of the expansion tensor, respectively. Therefore, we compute (see Appendix \ref{AppA} for details)
\beq
\theta_a{}^b \qquad\text{such that} \qquad \theta_a{}^b k_b=0 \qquad \theta_a{}^b q_{bc}=\theta_{ac}.
\eeq
These defining conditions translate in the exact-in-$\lambda$ expressions
\beqn
\theta_a{}^u  = 0 \qquad
\theta_a{}^B q_{Bc} =\theta_{ac} \label{thetaaB}.
\eeq
Then, we can evaluate the various components:
\beqn
\theta_a{}^u  &=& 0\\
\theta_A{}^B &=& \thetazero_A{}^B+\frac1{\lambda}\thetaone_A{}^B+\cO(\lambda^{-2})\\
\theta_u{}^A &=& -\frac1{\lambda^2}\ellzero^B\thetazero_B{}^A-\frac1{\lambda^3}\left(\ellone^B\thetazero_B{}^A+\ellzero^B\thetaone_B{}^A\right)+\cO(\lambda^{-4}),
\eeq
where 
\beq
\thetazero_A{}^B&=&\frac12 \pa_u \bq_{AC}\bq^{CB}\\
\thetaone_A{}^B&=&\frac12 N_A{}^B-\frac12 \pa_u q_{AC}C^{CB}-\bar F \delta_A^B
\eeq
and we introduced $N_A{}^B=N_{AC}\bq^{CB}$. 

Decomposing $\theta_a{}^b$ into its trace and traceless parts,
\beq
\theta_a{}^b=\frac{\theta}{2}q_a{}^b+\sigma_a{}^b,
\eeq
we find that the expansion is given by
\beq\label{theta}
\theta= \frac12 \bq^{AB}\pa_u \bq_{AB}-\frac2{\lambda}\bar F+\cO(\lambda^{-2})
\eeq
while the shear reads
\beq\label{sigmauA}
\sigma_a{}^u=0\qquad \sigma_u{}^A=\frac1{\lambda^2}\left(-\ellzero^B\thetazero_B{}^A+\frac14(\bar U^A+\bar q^{AB}\pa_B c_0) \bq^{CD}\pa_u \bq_{CD}\right)+\cO(\lambda^{-3}),
\eeq
and
\beq\label{sigmaAB}
\sigma_A{}^B=\frac12 \pa_u \bq_{AC}\bq^{CB}-\frac14 \delta_A^B \bq^{CD}\pa_u \bq_{CD}+\frac1{2\lambda}\left( N_A{}^B- \pa_u q_{AC}C^{CB}\right)+\cO(\lambda^{-2}).
\eeq
This is the first important result of this manuscript: The shear $\sigma_{a}{}^b$ possesses a leading order which is independent of the Bondi shear $C_{AB}$. The next order contains the News tensor, the radiative degrees of freedom, but it does not depend on $\bar F$, which is indeed a pressure-like term, and thus it contributes to the trace of the stress tensor. One can easily verify that the shear is traceless up to the desired order, $\sigma_a{}^a=\sigma_A{}^A=\cO(\lambda^{-2})$. The appearance of the News tensor in the expansion of the $u$-shear can already be anticipated in \eqref{exo}: the leading orders of this tensor are the $\pa_u$ derivatives of eq. \eqref{qdd}, in which the asymptotic shear $C_{AB}$ appears. Therefore, the shear of the $\ell$ congruence captures the News in the subleading term of its asymptotic development.

\subsection{Einstein Equations}\label{sec:einstein}

The Einstein equations projected to the null hypersurface can be written as the conservation laws of the Carrollian stress tensor. This is the null Brown-York stress tensor discussed in \cite{Chandrasekaran:2020wwn, Chandrasekaran:2021hxc, Chandrasekaran:2021vyu,  Freidel:2022bai, Ciambelli:2023mir}, which plays a prominent role in organizing the local degrees of freedom on a finite distance null hypersurface. It is defined as
\beq\label{T}
T_a{}^b=\frac1{8\pi G}\left(D_a\ell^b-\delta_a^b D_c\ell^c\right)=\tau_a\ell^b+\tau_a{}^b,
\eeq
with
\beq\label{ttt}
\tau_a=\frac1{8\pi G}(\pi_a-\theta k_a),\qquad \tau_a{}^b=\frac1{8\pi G}\left(\sigma_a{}^b-\mu q_a{}^b\right).
\eeq
On top of quantities already defined, in \eqref{ttt} we have introduced the surface tension $\mu=\kappa+\frac12 \theta$. This is a useful combination, as it leads to a canonical phase space variable \cite{Hopfmuller:2016scf, Hopfmuller:2018fni}, crucial in the quantization \cite{Ciambelli:2024swv}. Furthermore, the connection appearing in \eqref{T} is the Carrollian connection coming from the induced Rigging connection \cite{Mars:1993mj}. We define it and discuss its properties in appendix \ref{AppA}. 

Then, the constraints on the null hypersurface, that is, the projected Einstein equations, are given by \cite{Chandrasekaran:2020wwn, Chandrasekaran:2021hxc, Chandrasekaran:2021vyu,  Freidel:2022bai, Ciambelli:2023mir}
\beq\label{DTT}
8\pi G D_b T_a{}^b=0,
\eeq
for vacuum Einstein gravity. Our goal is to expand \eqref{DTT} asymptotically to retrieve the leading order Einstein equations. This in turns allows us to appreciate the asymptotic Einstein equations as conservation laws for the stress tensor $T_a{}^b$.

\subsubsection*{Raychaudhuri Equation}

The Raychaudhuri equation is the temporal component of the projection of Einstein's equations on the hypersurface. For vacuum Einstein equations, it reads
\beq\label{Ray}
8\pi G\,\ell^bD_aT_b{}^a=0 \quad \Rightarrow \quad \left(\cL_{\ell}+\theta\right)\theta=\mu \theta-\sigma_a{}^b \sigma_b{}^a.
\eeq
Using the asymptotic limit of all the quantities involved, \eqref{kappa}, \eqref{lu}, \eqref{lA}, (\ref{theta}-\ref{sigmaAB}), we find at leading order (that is, $\lambda^0$)
\beq\label{Ray1}\boxed{
\pa_u (\bq^{AB}\pa_u \bq_{AB})=-\frac12 \pa_u \bq_{AC}\bq^{CB} \pa_u \bq_{BD}\bq^{DA}.}
\eeq
The leading order Raychaudhuri equation constraints the boundary metric only. In other words, the time evolution of the boundary metric is constrained as an initial value problem, its evolution is not dictated by the radiation profile reaching asymptotic infinity. This is to be contrasted with the finite-distance case, where radiation sources the geometric structure of the null hypersurface. It is precisely this hierarchy of asymptotic equations that allows us to treat gravitons non-perturbatively at asymptotic null infinity, as they do not backreact on the boundary geometric data. This statement might be surprising at first, but it is the deep reason why the asymptotic quantization of gravity eludes the necessity to introduce quantum geometric operators, and reduces to creation and annihilation of gravitons, even in the strong gravity regime where Newton constant is large. This is arguably one of the profound reasons why the celestial holography program\footnote{See the reviews \cite{Strominger:2017zoo, Raclariu:2021zjz, Pasterski:2021rjz, Pasterski:2023ikd} and references therein.} is achieving successful results in understanding the structure of flat-space gravity from its null boundaries, by recasting it as a holographic field theory on a fixed background, see also discussions in \cite{Ciambelli:2024kre}.

\subsubsection*{Damour Equation}

The Damour equation is the horizontal part of the projection of Einstein's equations on the hypersurface. For vacuum Einstein equations, it reads
\beq\label{Damour}
8\pi G\,q_a{}^bD_cT_b{}^c
=0 \quad \Rightarrow \quad q_a{}^b \left({\cal L}_\ell+\theta\right)\pi_b+\theta\varphi_a=(\overline{D}_b+\varphi_b)(\mu q_a{}^b-\sigma_{a}{}^b),
\eeq
where $\overline{D}_a$ is the projected horizontal derivative defined in Appendix \ref{AppA}.

The quantity $\varphi_a$ appearing in \eqref{Damour} is the Carrollian acceleration, which is defined together with the Carrollian vorticity $w$ via
\beq\label{varphi}
\rd k=\varphi\wedge k+w, \quad \Rightarrow \quad \varphi_{a}=-\cL_{\ell}k_{a}.
\eeq
Its asymptotic expansion is
\beq\label{varphiuA}
\varphi_u=0,\qquad \varphi_A=\frac2{\lambda^2}\pa_A\bar\beta+\cO(\lambda^{-3}).
\eeq
We next evaluate the H\'aji\u{c}ek connection \eqref{defkappa} and find
\beq\label{pi}
\pi_u=0\qquad \pi_A=\frac1{\lambda}\left(\bar U_A+\pa_A c_0\right)+\frac1{\lambda^2}\left(\pa_A c_1-c_0(\bar U_A+\pa_A c_0)\right)+\cO(\lambda^{-3}).
\eeq

To compute the horizontal covariant derivative, one must compute in an adapted frame the Christoffell symbols. However, we remark that for the first two orders in the $\lambda$ expansion of all the quantities involved in the right-hand side of \eqref{Damour}, only the spatial components survive. Therefore, if we truncate the expansion to second order, we can safely replace
\beq
\overline{D}_b (\mu q_a{}^b-\sigma_a{}^b)=\delta_a^A \overline{D}_B(\mu q_A{}^B-\sigma_A{}^B)+\text{subleading orders in} \ \lambda.
\eeq
This is useful because the covariant derivative $\overline{D}_A$ is simply now the Levi-Civita derivative of the non-degenerate metric given by the spatial part of $q_{ab}$ in \eqref{qdd}, that is,
\beq
q_{AB}=\lambda^2 \bq_{AB}+\lambda(C_{AB}+2c_0 \bq_{AB}).
\eeq

We want to study the leading order in $\lambda$ of \eqref{Damour}. Since $\pi_a$ and $\varphi_a$ are both subleading with respect to $\theta$ and $\sigma_a{}^b$, we gather 
\beq
\oD_B (\mu q_A{}^B-\sigma_A{}^B)=0.
\eeq
Using then that $\kappa$ is subleading with respect to $\theta$, and \eqref{qdU}, we obtain
\beq
\oN_B \left(\frac{\theta}{2} \delta_A^B-\sigma_A{}^B\right)=0,
\eeq
where to first order $\oD_B$ is simply the covariant derivative of the boundary metric $\bq_{AB}$, called $\oN_B$. Combining \eqref{theta} and \eqref{sigmaAB}, we process
\beq
\frac{\theta}{2} \delta_A^B -\sigma_A{}^B=\frac14\delta_A^B \bq^{CD}\pa_u \bq_{CD}-\frac12 \pa_u \bq_{AC}\bq^{CB}+\cO(\lambda^{-1}),
\eeq
such that the Damour equation at leading asymptotic order is
\beq\label{Dam1}\boxed{
\oN_A (\pa_u \bq_{CD}\bq^{CD})-\oN_B(\pa_u \bq_{AC}\bq^{CB})=0.}
\eeq

With this equation, we conclude the general asymptotic analysis. We obtained the leading order expressions for the Raychaudhuri equation and Damour equation, \eqref{Ray1} and \eqref{Dam1} respectively.  These equations are readily solved by \eqref{dq}. We stress again that we have suppressed the order $r$ term in \eqref{fo1}, in the expansion of $F$, see discussion below \ref{eom}. If one allows for such a term, then the leading equations of motion constraint the boundary metric to be conformally time independent, as originally derived in \cite{Barnich:2010eb}. This is important to understand the Weyl structure arising at the boundary. However, we will not pursue that road, but rather the opposite. That is, we will work in a simplified scenario where the boundary metric is completely time independent. Indeed, without making further simplifying assumptions on the spatial metric at the boundary, going into subleading orders is technically complex, and not particularly enlightening.

\section{Simplified Framework}\label{sec:simplified}

While we could continue and explore the subleading expressions, this quickly becomes an untractable exercise. Instead, leveraging that both Raychaudhuri and Damour asymptotic equations involve  $\bq_{AB}$ only, and thus constraining the boundary metric does not restrict the radiative solution space, we solve these equations imposing \eqref{dq}, that is,
\beq\label{dqe0}
\pa_u \bq_{AB}=0.
\eeq
This allows us to make further contact with most of the existing literature. To simplify our analysis even more, we require the metric of the cut to be 
\beq
\rd s_{\mathrm{cut}}=2 \bq_{z\bz}(z,\bz) \rd z \rd \bz\,,
\eeq
such that we have a non-constant curvature scalar yet we automatically satisfy the leading order constraints.
Repeating the same steps as in the previous sections, we can then extract the leading order equations of motions in the simplified setup \eqref{dqe0}. The derivations presented in sections \ref{sec:constructing} and \ref{sec:asymptotic} are unchanged. The first important consequence of \eqref{dqe0} for the boundary metric occurs in section \ref{sec:intrinsic}, when computing the expansion tensor.

\subsection{Intrinsic Carrollian Data}\label{sec:intrinsic2}

The geometric data on the asymptotic null hypersurface are the Carrollian vector field \eqref{lu} and \eqref{lA}, the Ehresmann connection \eqref{ku}, and the induced degenerate metric \eqref{qdd}, with $\bq_{AB}$ given by
$\bq_{zz}=0=\bq_{\bz\bz}$ and $\pa_u \bq_{z\bz}=0$. To get the Bondi mass-loss formula we must expand one order more in $q_{AB}$. Doing so we get
\beq\label{qsuu}
q_{uu}  & =& \cO(\lambda^{-2}),\\
q_{u A} & =& -\bar{U}_A-\pa_A c_0 - \frac{1}{\lambda}\left( C_{AB}\bar{U}^B+U_{1A}+\pa_A c_1\right)+\cO(\lambda^{-2}),\\
q_{AB} & =& \lambda^{2} \bar q_{AB} + \lambda (C_{AB}+2 c_0 \bar q_{AB})+q^{(0)}_{AB}  + O(\lambda^{-1}),\label{qsAB}
\eeq
where 
\beq
q^{(0)}_{AB}=c_0 C_{AB}+\left(c_0^2+2c_1+\frac{C_{CD}C^{CD}}{4}\right)\bq_{AB}.
\eeq

The asymptotic expansion tensor significantly changes. Recalling its definition,
\beq
\theta_{ab}=\frac12 \cL_{\ell}q_{ab},
\eeq
its components become 
\beq
\theta_{AB}&=&\lambda\left(\frac12 N_{AB}-\bar F \bq_{AB}\right)+\theta^{(0)}_{AB}+\cO(\lambda^{-1})\\
\theta_{uA}&=&-\frac{\ellzero^B}{\lambda}\left(\frac12 N_{AB}-\bar F \bq_{AB}\right)+\cO(\lambda^{-2}) \\
\theta_{uu}&=&\cO(\lambda^{-2}),
\eeq
where again $N_{AB}=\pa_u C_{AB}$ is the News tensor (now automatically traceless), and we introduced
\beq
\theta^{(0)}_{AB}=\frac12 \pa_u q^{(0)}_{AB}+\frac12 \bar\cL_{\ellzero}\bq_{AB},
\eeq
where by $\bar\cL_{\ellzero}\bq_{AB}$ we mean the spatial Lie derivative with respect to $\ellzero^A$, which we  recall is given by $\ellzero^A=\bar U^A+\bar q^{AB}\pa_B c_0$.

By design, $\theta_{ab}$ is orthogonal to $\ell^a$,
\beq
\ell^a\theta_{ab}=\cO(\lambda^{-2}).
\eeq
Clearly, in our simplified setup the leading order in \eqref{theta0} vanishes, and thus we have access to subleading contributions, where radiative degrees of freedom reside.

The next step is to evaluate
\beq
\theta_a{}^b \qquad\text{such that} \qquad \theta_a{}^b k_b=0 \qquad \theta_a{}^b q_{bc}=\theta_{ac}.
\eeq
While $\theta_a{}^u=0$, the non-vanishing components are simply
\beqn\label{tdU}
\theta_A{}^B=\frac1{2\lambda}(N_A{}^B-2\bar F \delta_A^B)+\frac1{\lambda^2}(\theta^{(0)}_A{}^B-(\frac{N_{AC}}{2}-\bar F \bq_{AC})(C^{CB}+2c_0 \bq^{CB}))+\cO(\lambda^{-3})\,\,\,\,\,
\eeqn
and
\beqn
\theta_u{}^A = -\frac{\ellzero^B}{\lambda^3}\left(\frac12 N_B{}^A-\bar F \delta_B^A\right)+\cO(\lambda^{-4})\,,
\eeq
where now there are no subtleties in $N_A{}^B=\bq^{BC}\pa_u C_{AC}=\pa_u C_A{}^B$, since the metric of the cut is time-independent. It is instructive to process the order-$\lambda^{-2}$ term in \eqref{tdU}. Using \eqref{conds}, it gives
\beqn\label{th0}
\theta^{(0)}_A{}^B-(\frac{N_{AC}}{2}-\bar F \bq_{AC})(C^{CB}+2c_0 \bq^{CB})&=&\frac12 \bar\cL_{\ellzero}\bq_{AC} \bq^{CB}-\frac{N_{AC}C^{CB}}{2}-\frac{c_0 N_A{}^B}{2}\\
&&+\frac{\bar F C_A{}^B}{2}+\delta_A^B\left(c_0 \bar F+m+\frac{N_{CD}C^{CD}}{4}\right)\,. \nonumber
\eeqn
We need this term because we need the leading and first subleading terms in the expansion scalar $\theta$. Indeed, one remarks that the Bondi mass appears in \eqref{th0} multiplying $\delta_A^B$, and thus it is a pure trace contribution.

The decomposition $\theta_a{}^b=\frac{\theta}{2}q_a{}^b+\sigma_a{}^b$ gives the expansion scalar
\beq\label{the}
\theta=-\frac2{\lambda}\bar F+\frac1{\lambda^2}\left(2 c_0 \bar F+2 m+\oN_A\ellzero^A\right)+\cO(\lambda^{-3})\,,
\eeq
where we used $\frac12 \bar\cL_{\ellzero}\bq_{AC} \bq^{CA}=\oN_A\ellzero^A$, while the shear reads
\beq\label{suA}
\sigma_a{}^u=0\qquad \sigma_u{}^A=-\frac{\ellzero^BN_B{}^A}{2\lambda^3}+\cO(\lambda^{-4})
\eeq
and
\beq\label{sAB}
\sigma_A{}^B=\frac{N_A{}^B}{2\lambda}+\frac1{\lambda^2}\sigma_A^{(2)B}+\cO(\lambda^{-3}),\qquad
\eeq
with
\beq
\sigma_A^{(2)B}=\frac1{4}(C_{AC}N^{CB}-N_{AC}C^{CB}+2(\bar F C_A{}^B-c_0 N_A{}^B- \bar\cL_{\ellzero}\bq_{\langle A}{}^{B\rangle}))
\eeq
where $\bar\cL_{\ellzero}\bq_{\langle A}{}^{B\rangle}$ denotes the traceless part of $\bar\cL_{\ellzero}\bq_{AC} \bq^{CB}$.
The shear contains indeed the information on the radiation. On the other hand, the expansion is controlled by $\bar F$ at leading order, and by $c_0$, $m$, and the spatial expansion of $\ellzero$ at subleading order. The finite distance expansion is the Hamiltonian charge associated to null translation. From the shape of its asymptotic expansion we learn that the boundary curvature and the Bondi mass play the role of Hamiltonian charge densities for the null evolution on the boundary. While the soft sector of the theory has been recently discussed in \cite{He:2024vlp}, it would be interesting to perform a complete matching of asymptotic charges from a bulk null hypersurface.

\subsection{Einstein Equations}\label{sec:einstein2}

We now construct the asymptotic Raychaudhuri and Damour equations in the simpler setting \eqref{dqe0}. Calling $C$ the Raychaudhuri constraint in \eqref{Ray},
\beq
C=\left(\cL_{\ell}+\theta\right)\theta-\mu \theta+\sigma_a{}^b \sigma_b{}^a,
\eeq
we can expand in powers of $\lambda$. We find
\beq
C=C_{(0)}+\frac1{\lambda}C_{(1)}+\frac1{\lambda^2}C_{(2)}+\cO(\lambda^{-3}).
\eeq
We already computed the leading order in \eqref{Ray1},
\beq
C_{(0)}=\pa_u (\bq^{AB}\pa_u \bq_{AB})+\frac12 \pa_u \bq_{AC}\bq^{CB} \pa_u \bq_{BD}\bq^{DA},
\eeq
which here is automatically satisfied as $\pa_u\bq_{AB}=0$. Imposing this and continuing, we gather
\beq\label{c11}
C_{(1)}=-2 \pa_u \bar F.
\eeq
We kept $\bar F$ free on purpose in our computations to arrive to this result. Eq. \eqref{c11} is perfectly compatible with the leading asymptotic Einstein equations normal to the hypersurface, that we collected in \eqref{eom}, and recall here:
\beq
\bar\beta+\frac1{32}C_{AB}C^{AB}=0\qquad \bar R=4\bar F\qquad \bar U^A+\frac12  \oN_B C^{AB}=0.
\eeq
Therefore, using that the boundary metric is time-independent, and substituting in our computations $\bar F=\frac{\bar R}{4}$, we automatically satisfy $C_{(1)}=0$. Imposing as well $ \bar U^A=-\frac12  \oN_B C^{AB}$, we then reach
\beq\boxed{\label{C2}
C_{(2)}=2\left(\pa_u m-\frac14 \oN_A \oN_B N^{AB}-\frac18 \bar \Delta \bar R+\frac18 N_{AB}N^{AB}\right),
}\eeq
which is exactly the Bondi mass-loss formula \eqref{bmlf}. Eq. \eqref{C2} is one of the main results of this manuscript. We have shown that the Raychaudhuri equation for a null asymptotic hypersurface gives rise to the Bondi mass-loss formula. While this is expected from first principle, based on the fact that these are both the null temporal components of Einstein equations, understanding the Bondi mass-loss formula as a Raychaudhuri constraint unlocks a plethora of key consequences. We will explore two such consequences in the following, namely, the matching of the asymptotic phase space and the construction of an asymptotic stress tensor. It is moreover our intention to investigate this result in depth in future works. One important aspect is that the Raychaudhuri constraint has been understood as a Carrollian conservation law of a stress tensor at finite distance, while such a result is missing for the Bondi mass loss formula. This paper is filling this gap. Other important repercussions of this result concern the quantization addressed in \cite{Ciambelli:2024swv}, and its fate in the asymptotic limit. With the results of this manuscript, the stage is set to pursue this analysis and relate finite distance results to celestial and Carrollian flat-space holography.

Let us now turn our attention to the Damour equation \eqref{Damour}. Calling $J_a$ the Damour  constraint,
\beq\label{Ja}
J_a=q_a{}^b \left({\cal L}_\ell+\theta\right)\pi_b+\theta\varphi_a-(\overline{D}_b+\varphi_b)(\mu q_a{}^b-\sigma_{a}{}^b),
\eeq
we have the asymptotic expansion
\beq
J_a=J^{(0)}_a+\frac1{\lambda}J^{(1)}_a+\frac1{\lambda^2}J^{(2)}_a+\cO(\lambda^{-3}).
\eeq
The leading order is given by \eqref{Dam1}, which is automatically solved here. We then evaluate the subleading order and find
\beq\label{j1}
J^{(1)}_u=0,\qquad J^{(1)}_A=\pa_u\bar U_A+\frac12  \oN_B N_{A}{}^{B}.
\eeq
Similar to the Raychaudhuri equation, the first subleading Damour constraint gives an equation fully compatible with the asymptotic Einstein equations normal to the hypersurface. Indeed, eq. \eqref{j1} is simply the $u$ derivative of the last equation in \eqref{eom}. 

Imposing \eqref{j1} and continuing, we then move to the order $\lambda^{-2}$. We first focus on $J^{(2)}_u$. From \eqref{quA}, \eqref{varphiuA},  \eqref{pi}, \eqref{suA}, and the fact that $\mu=\cO(\lambda^{-1})$, we readily obtain that $J^{(2)}_u=0$, since the first non-trivial contribution to this component is at order $\lambda^{-3}$. Concerning $J^{(2)}_A$, the computations involved become quickly untractable. We will not pursue this computation here, but we stress that it is on our agenda. This is important in order to demonstrate that the angular-momentum equation of motion \eqref{pauP} is encrypted in the Damour equation. This however requires to go to very subleading orders, as $\bar P_A$ appears in $\Uone_A$, which is expected to contribute at best at order $\lambda^{-3}$. 

\subsection{Holographic Stress Tensor}\label{sec:holographic}

We recall that the null Brown-York stress tensor on a null hypersurface is given by
\beq
T_a{}^b=\tau_a\ell^b+\tau_a{}^b,\qquad 
\tau_a=\frac1{8\pi G}(\pi_a-\theta k_a),\qquad \tau_a{}^b=\frac1{8\pi G}\left(\sigma_a{}^b-\mu q_a{}^b\right).
\eeq
We are interested in its asymptotic expansion, and mostly in its first non-vanishing terms.
Using equations \eqref{ku}, \eqref{pi}, and \eqref{the}, we get
\beqn\label{tu}
\tau_u&=&\frac{\bar F}{4\pi G \lambda}-\frac1{8\pi G \lambda^2}\left(2 c_0 \bar F+2 m+\oN_A\ellzero^A\right)+\cO(\lambda^{-3})\\
\tau_A&=&\frac1{8\pi G\lambda}\ellzero_A+\frac1{8\pi G\lambda^2}\left(\pa_A c_1-c_0\ellzero_A\right)+\cO(\lambda^{-3})\,,\label{tA}
\eeqn
where $\ellzero_A=\bar q_{AB}\ellzero^B$.
Similarly, from \eqref{kappa}, (\ref{the}-\ref{sAB}), and (\ref{qdU},\ref{quA}), we gather
\beqn
\tau_u{}^u&=&0\\
\tau_A{}^u&=&0\\
\tau_u{}^A&=&-\frac{1}{16\pi G\lambda^3}\left(\ellzero^B N_B{}^A+2\bar F\ellzero^A\right)+\cO(\lambda^{-4})\\
\tau_A{}^B&=&\frac1{16\pi G\lambda}\left(N_A{}^B+2\bar F \delta_A^B\right)\\
&&+\frac1{16\pi G\lambda^2}\left(2\sigma_A^{(2)B}-(2c_0 \bar F+2m+\oN_C\ellzero^C)\delta_A^B\right)+\cO(\lambda^{-3})\,.\label{tAB}
\eeqn

These results, together with (\ref{lu},\ref{lA}), allow us to finally express the asymptotic limit of the null Brown-York stress tensor
\beqn
T_u{}^u&=&\frac{\bar F}{4\pi G \lambda}-\frac1{8\pi G \lambda^2}\left(2 c_0 \bar F+2 m+\oN_A\ellzero^A\right)+\cO(\lambda^{-3})\\
T_A{}^u&=&\frac1{8\pi G\lambda}\ellzero_A+\frac1{8\pi G\lambda^2}\left(\pa_A c_1-c_0\ellzero_A\right)+\cO(\lambda^{-3})\\
T_u{}^A&=&-\frac{1}{16\pi G\lambda^3}\left(\ellzero^B N_B{}^A-2\bar F\ellzero^A\right)+\cO(\lambda^{-4})\\
T_A{}^B&=&\frac1{16\pi G\lambda}\left(N_A{}^B+2\bar F \delta_A^B\right)\\
&&+\frac1{16\pi G\lambda^2}\left(2\sigma_A^{(2)B}-(2c_0 \bar F+2m+\oN_C\ellzero^C)\delta^B_A\right)+\cO(\lambda^{-3})\,.
\eeqn
This is one of the main results of this paper. As much as the Balasubramanian-Kraus stress tensor plays a crucial role in AdS/CFT \cite{brown1993quasilocal, Balasubramanian:1999re, Emparan:1999pm, deHaro:2000vlm}, we expect that this stress tensor will play an important role in flat space holography, and its relationship to the various proposals in the celestial \cite{Kapec:2016jld, Kapec:2017gsg} and Carrollian \cite{Bagchi:2015wna, Ciambelli:2018wre, Saha:2023hsl, Adami:2023fbm, Saha:2023abr, Ciambelli:2024kre, Ruzziconi:2024kzo, Adami:2024rkr, Bagchi:2024gnn} literature is part of our agenda. Note that the boundary curvature and Bondi mass appear in the time-time component. Indeed, they give rise to the asymptotic energy of the system. This is the reason why they dictate the position of an asymptotic null observer, see \eqref{pos}. In \cite{Riello:2024uvs, Bhambure:2024ftz}, a stress tensor for flat space has been proposed along similar lines of the one introduced here. The main -- and crucial -- difference is that in these references the bulk hypersurface is timelike (the stretched horizon), whereas we are here considering a family of null hypersurfaces. This implies that our stress tensor and theirs slightly differ. Technically, this arises because the expansion considered there is in powers of the radial coordinate $\Omega=\frac{1}{r}$, which is null only in the asymptotic limit, whereas our expansion ensures the hypersurface to remain null as we enter the bulk. The reason why we chose to perform this analysis with a family of null hypersurfaces is to maintain the nature of the hypersurface in the family, such that null infinity is not geometrically special. Then, tools such as those employed in \cite{Ciambelli:2023mir} can be exported to null infinity straightforwardly.

We can take here is a simplified setup, where we Weyl-rescale the spatial part of the boundary to be flat space, $\bar R=0$. In this simple framework, the stress tensor has leading orders:
\beq\boxed{
T_a{}^b=\frac1{16\pi G \lambda}\begin{pmatrix}
0 & 0\\
 2\ellzero_A & N_A{}^B
\end{pmatrix}+\frac1{8\pi G\lambda^2}\begin{pmatrix}
-2 m -\oN_C\ellzero^C& 0\\
\pa_A c_1-c_0 \ellzero_A & \sigma_A^{(2)B}-(m+\frac12 \oN_C\ellzero^C)\delta_A^B
\end{pmatrix}+\cO(\lambda^{-3})\nonumber}
\eeq
where we recall that $\pa_u c_1=m$.

This allows us also to provide a preliminary fluid interpretation of the boundary data: the Bondi mass plays the role of the energy of the system, whereas the Bondi News is the viscous shear. Interestingly, there is a current-like term, $\pa_A c_1-c_0 \ellzero_A$, which depends on the Bondi mass profile everywhere on $\cN$. Speculatively, this term could be associated to a heat current, and thus $c_1$ to a notion of temperature. It would be rewarding to pursue this fluid interpretation, which is the thread connecting \cite{Ciambelli:2018wre} (see also \cite{Ciambelli:2018ojf}) with \cite{Ciambelli:2023mir}.

If Witten dictionary \cite{Witten:1998qj} pertains to flat-space holography, this stress tensor would be interpreted as the response of the boundary Carrollian system under a perturbation (source) generated by varying the boundary data. The latter could either be the boundary metric and Carrollian vector field $\ell$, or, as the asymptotic phase space suggests, the asymptotic shear $C_{AB}$. We plan to study this stress tensor in details in the future, and to relate it to the $S$-matrix analysis of \cite{Kraus:2024gso}.

We conclude this subsection addressing an important question: where does the angular momentum aspect appear in the stress tensor? As we anticipated at the end of section \ref{sec:einstein2}, this quantity appears at order $\lambda^{-3}$, and thus subleadingly to the orders displayed in the stress tensor. There are two reasons for this fact. The first reason is that the induced Carrollian connection has an expansion in $\lambda$ which mixes the $\lambda$-terms in the stress tensor when taking the divergence to construct the equations of motion. This does not happen only for the leading order, for which  the asymptotic temporal equation of motion is indeed the mass conservation law. Nonetheless, already at this order, since the Christoffel symbols depend on $\lambda$, various orders of the stress tensor components may contribute. Let us show this explicitly. Considering the stress tensor, we wish to demonstrate that the Bondi mass-loss formula is the temporal contraction of the covariant derivative of the asymptotic null Brown-York stress tensor
\begin{equation}\label{ldt}
    \ell^a D_b T_a{}^b=D_b T_u{}^b=-\frac1{8\pi G\lambda^2}\pa_u(2 m +\oN_C\ellzero^C)-\Gamma_{bu}^aT_a{}^b+\Gamma_{ba}^b T_u{}^a+\cO(\lambda^{-3})\,.
\end{equation}
The leading orders of $\Gamma_{bu}^a$ are given by
\begin{equation}\label{Gamma}
\Gamma_{bu}^a=\frac1{2\lambda}N_A{}^B \delta^A_b\delta_B^b+{\cal O}(\lambda^{-2}) \qquad \Gamma_{au}^a=\frac1{2\lambda}N_A{}^A+{\cal O}(\lambda^{-2})={\cal O}(\lambda^{-2}),
\end{equation}
which implies that the $1/\lambda^2$ piece in the $T_u{}^u$ component mixes with the $1/\lambda$ piece in the $T_A{}^B$ component, because $\Gamma_{bu}^a$ starts at order $1/\lambda$.

Therefore, eq.~\eqref{ldt}, to leading order in $\lambda$, is
\beq
    \ell^a D_b T_a{}^b&=&-\frac1{8\pi G\lambda^2}\pa_u(2 m +\oN_C\ellzero^C)-\frac1{2\lambda}N_A{}^BT_B{}^A+\cO(\lambda^{-3})\\
    &=&-\frac1{4\pi G\lambda^2}\left(\partial_u m+\frac12\oN_C\pa_u\ellzero^C+\frac18 N_A{}^BN_B{}^A\right)+\cO(\lambda^{-3})\,.
\eeq
Then, using \eqref{eom}, we obtain
\beq
    \ell^a D_b T_a{}^b=-\frac1{4\pi G\lambda^2}\left(\partial_u m-\frac14\oN_C\oN_B N^{BC}+\frac18 N_A{}^BN_B{}^A\right)+\cO(\lambda^{-3})\,,
\eeq
which is exactly eq. \eqref{bmlf} with $\bar R=0$. We thus see that, as soon as we go into the subleading orders, the non-trivial expansion of the connection mixes the orders in the stress tensor. This is the first reason why the angular momentum and the mass aspects appear at different orders. 

The second reason is that we are here working on the physical spacetime. A preliminary result (this is currently under investigation) suggests that indeed the orders in the $\lambda$-expansion mixes differently when considering the conformally-compactified spacetime. It is important to pursue this computation, to be able to compare with the existing literature \cite{Riello:2024uvs, Bhambure:2024ftz}, which deals with the unphysical spacetime only. In this upcoming work, we will be in the same framework as these papers, and so we will offer a scrupulous comparison.

\subsection{Covariant Phase Space}\label{sec:covariant}

The gravitational covariant phase space induced on a finite-distance null hypersurface has been studied by many authors \cite{Hayward:1993my,  Reisenberger:2007pq, Lehner:2016vdi, Donnay:2016ejv, Wieland:2017zkf, Hopfmuller:2018fni, Chandrasekaran:2018aop, Donnay:2019jiz, Chandrasekaran:2020wwn, Adami:2021nnf, Chandrasekaran:2021hxc, Chandrasekaran:2021vyu, Sheikh-Jabbari:2022mqi, Odak:2023pga, Ciambelli:2023mir}. We utilize here the formulation and framework of \cite{Chandrasekaran:2020wwn, Chandrasekaran:2021hxc, Ciambelli:2023mir}. The pre-symplectic potential is 
\beq\label{thcan}
\theta_{\mathrm{can}}=\int_{\cN}\ve_{\cN} \left(\frac12 \tau^{ab}\delta q_{ab}-\tau_a\delta \ell^a\right)\,.
\eeq

Our goal is to perform the asymptotic expansion of this phase space. From (\ref{qsuu}-\ref{qsAB}), we get
\beq
\delta q_{ab}= \delta_a^A\delta_b^B  \left(\lambda^2 \delta \bar q_{AB}+\lambda (\delta C_{AB} +2 \delta (c_0 \bar q_{AB}) \right) +\cO(\lambda^0).
\eeq
While it is certainly interesting to keep the leading order boundary metric dynamical in the phase space,\footnote{See \cite{Compere:2019bua, Compere:2020lrt, Geiller:2022vto, Geiller:2024amx, Campiglia:2024uqq} for recent analyses in this direction.} we here assume $\delta \bar q_{AB}=0$, to make contact with the existing literature. We further require $\delta c_0=0$, which implies we are not letting the position of the null hypersurface fluctuate in our phase space. Relaxing this would be also an extension of this work worth pursuing. We then gather
\beq
\delta q_{ab}= \delta_a^A\delta_b^B \lambda \delta C_{AB}+\cO(\lambda^0).
\eeq
On the other hand, from (\ref{lu}-\ref{lA}), we get
\beq
\delta \ell^a=\cO(\lambda^{-2}).
\eeq
Then, given \eqref{tu} and \eqref{tA}, we get that the spin-$1$ contribution $\tau_a\delta \ell^a$ to \eqref{thcan} is subleading with respect to the spin-$2$ and spin-$0$ contributions $\tau^{ab}\delta q_{ab}$.

Using \eqref{tAB}, to leading asymptotic order we get\footnote{From \eqref{qsAB}, the leading order inverse metric in the spatial directions is simply $\frac1{\lambda^2} \bar q^{AB}$.}
\beq
\tau^{AB}=\frac{1}{16\pi G \lambda^3}\left(N^{AB}+2\bar F \bar q^{AB}\right)+\cO(\lambda^{-4}).
\eeq
Then, using that $C_{AB}$ is traceless and thus $\bar q^{AB} \delta C_{AB}=0$, we collect
\beq
\theta_{\mathrm{can}}&=&\int_{\cN} \ve_{\cN} \left(\frac12 \tau^{ab}\delta q_{ab}-\tau_a\delta\ell^a\right)\\
&=&\frac12 \int_{\cN} \ve_{\cN}\tau^{AB} \left(\lambda \delta C_{AB}+\cO(\lambda^0)\right)\\
&=&\frac1{32\pi G\lambda^2} \int_{\cN} \ve_{\cN} \left(N^{AB} \delta C_{AB}+\cO(\lambda^{-1})\right).\label{thcanp}
\eeq

We now process the volume element, which at finite distance is given by
\beq
\ve_{\cN}=k\wedge \ve_{\cC}=\sqrt{q}\,\ve^{(0)}_{AB}\, k\wedge \rd \sigma^A \wedge\rd \sigma^B\,,
\eeq
where $\cC$ is a constant-$u$ cut and $\ve^{(0)}_{AB}$ is the two-dimensional Levi-Civita symbol. We now compute its asymptotic expansion. From \eqref{qsAB} we have
\beq
\sqrt{q}=\lambda^2 \sqrt{\bar q}+\cO(\lambda)\,,
\eeq
such that, using \eqref{ku},
\beq\label{ve}
\ve_{\cN}=\lambda^2 \sqrt{\bar q} \, \ve^{(0)}_{AB}\, \rd u \wedge \rd \sigma^A \wedge\rd \sigma^B +\cO(\lambda)=\lambda^2 \, \rd u\wedge \rd^2\Omega +\cO(\lambda)\,,
\eeq
where we introduced the infinitesimal volume form of the asymptotic constant-$u$ cut $\rd^2\Omega=\sqrt{\bar q} \, \ve^{(0)}_{AB}\, \rd \sigma^A \wedge \rd \sigma^B$.

Plugging \eqref{ve} into \eqref{thcanp}, we get 
\beq
\boxed{\theta_{\mathrm{can}}=\frac1{32\pi G} \int_{\cN} \rd u \, \rd^2\Omega\, N^{AB} \delta C_{AB}+\cO(\lambda^{-1}).}
\eeq
This leads precisely to the Ashtekar-Streubel gravitational phase space $\Omega_{\mathrm{AS}}=\delta\theta_{\mathrm{AS}}$ at null infinity \cite{Ashtekar1981} (see also \cite{Barnich:2010eb, Compere:2018ylh, Freidel:2021fxf, Ashtekar:2024stm}), numerical factors included. Indeed, we have shown that the asymptotic limit of the symplectic structure $\Omega_{\mathrm{can}}=\delta \theta_{\mathrm{can}}$ gives
\beq
\lim_{\lambda\to \infty}\Omega_{\mathrm{can}}=\Omega_{\mathrm{AS}}.\label{limlim}
\eeq

This is the main result of this section. As per the matching of asymptotic equations of motion, \eqref{limlim} is also expected. Nonetheless, this allows us to link the analysis of finite-distance hypersurfaces to null infinity. Many questions are now well-posed, and ready to be addressed. We collect them in the Conclusions hereafter. 

\section{Final Words}\label{sec:final}

In this work, we paved the way to the matching between physics on finite-distance null hypersurfaces and asymptotic null infinity. Although the BS gauge is not the most suitable to describe a family of null hypersurfaces parallel to null infinity, we used this gauge to make the comparison with existing literature more straightforward. We first studied how to describe a family of null hypersurfaces and their asymptotic limit. Then, we recast the induced geometric data from the bulk as intrinsic Carrollian quantities on the null hypersurfaces. Eventually, we studied the intrinsic Einstein constraints on these surfaces, the Raychaudhuri and Damour constraints, and show that they asymptote to the Einstein equations of motion for the boundary metric. We solved these leading equations requiring that the boundary metric is time-independent. This allowed us to probe the subleading structure of these equations, unveiling that the sub-sub-leading Raychaudhuri equation is exactly the Bondi mass-loss formula. 

Recasting asymptotically null physics in terms of the Raychaudhuri equation and finite-distance Carrollian physics is an important step. Indeed, we understood the Bondi mass-loss formula as the time component of the conservation law of the null Brown-York stress tensor. The asymptotic limit of the null Brown-York stress tensor is a new result of this manuscript. We concluded showing how the gravitational phase space induced on a finite-distance null hypersurface asymptote to the Ashtekar-Streubel phase space, linking therefore the finite-distance and asymptotic Noetherian analysis. 

While some of them have already been discussed throughout the manuscript, we recollect here a list of future directions we intend to explore. We begin with the more practical and computational ones.
\begin{itemize}
    \item \textit{Damour equation and angular momentum:} We wish to push the computation of the Damour asymptotic equation to subleading orders, in order to demonstrate that it leads to the angular momentum equation \eqref{pauP}. In this regard, it seems that the asymptotic analysis in the conformally-compactified spacetime mixes the orders in $\lambda$ differently, and thus could make the angular momentum appearing at more leading orders, facilitating furthermore the comparison with the existing literature.
    \item \textit{More general boundary conditions:} We want to study how to relax the asymptotic expansion of $F$, \eqref{fo1}, in order to allow for a time-dependent conformal factor in the boundary metric, as already touched upon below \eqref{eom}. This in turns allows us to fully appreciate the Weyl structure at null infinity.
    \item \textit{Relax phase space:} We have all the tools to make the boundary metric a dynamical variable on the phase space, $\delta\bar q_{AB}\neq 0$. We wish to explore this direction, comparing with the recent phase space analysis of \cite{Campiglia:2024uqq}.
    \item \textit{Corner terms:} We focused here on the bulk of the hypersurface $\cN$. What happens at the tips? We intend to study corner terms in the symplectic structure, and to match with those needed in the phase space  at asymptotic infinity. This could in turn highlights their relevance for the finite-distance phase space. 
\end{itemize}

On top of these technical questions, there are some more long-term directions unveiled by this work, that we are planning to investigate.
\begin{itemize}
    \item \textit{Fluid interpretation:} As much as the fluid/gravity correspondence \cite{Bhattacharyya:2007vjd} has been useful to understand the filling-in problem and the macroscopic aspects of the AdS/CFT duality, we expect the asymptotically flat fluid/gravity correspondence \cite{Ciambelli:2018wre, Campoleoni:2018ltl} to be a guiding light in flat-space holography. In the latter, a Carrollian stress tensor describing the boundary degrees of freedom is still lacking. Our proposal, together with the proposal described in \cite{Riello:2024uvs, Bhambure:2024ftz}, can provide a holographic stress tensor for flat space. In particular, we wish to investigate its hydrodynamic properties, and see if we can determine its transport coefficients. Incidentally, this can be relevant in setting up numerical GR analysis of the radiative bulk data in the presence of black holes.
    \item \textit{Charges and soft theorems:} One of the original motivations of this work is to see whether by entering into the bulk preserving the null nature of infinity one can learn more about the structure of asymptotic fluxes and charges, and their link to soft theorems. The latter have been related to symmetries in \cite{He:2014laa, Cachazo:2014fwa, Campiglia:2014yka}. Since here we have access to the subleading terms in the asymptotic expansion of the symplectic potential, \eqref{thcan}, we wish to study these terms and their relationship to asymptotic charges, fluxes, and symmetries of the $S$ matrix. 
    \item \textit{Bringing quantum information to flat-space holography:} Finite-distance null hypersurfaces have been the subject of intense studies from the quantum informational perspective. From the proof of the Generalized Second Law on horizons \cite{Wall:2011hj} and the Quantum Focusing Conjecture \cite{Bousso:2015mna}, to quantum energy bounds \cite{Bousso:1999xy, Hartman:2016lgu, Balakrishnan:2017bjg, Casini:2017roe, Ceyhan:2018zfg}, there is by now a tremendous amount of work on quantum informational properties of null hypersurfaces (see also \cite{Kontou:2020bta} and references therein). However, the application of these tools to flat-space has been so far elusive and scattered.\footnote{See, however, \cite{Li:2010dr, Apolo:2020bld, Rignon-Bret:2024zhj}, for a list of interesting works in this direction.} Our formalism could offer a fresh perspective on this topic, elevating the role of quantum information in flat-space holography to a level of prominence comparable to its position in AdS/CFT \cite{Ryu:2006bv, VanRaamsdonk:2010pw, Lewkowycz:2013nqa, Dong:2016eik}. 
    \item \textit{Phase space quantization:} In \cite{Ciambelli:2024swv}, we proposed a quantization of the phase space of gravity on a finite-distance hypersurface. In particular, we promoted the Raycahudhuri constraint to a quantum operator. This implied that the area is as well an operator. Based on \cite{Kapec:2016aqd}, we then reproduced in the asymptotic limit the infinite fluctuation of the Bondi mass found in \cite{Bousso:2017xyo}. The present work offers a framework to further pursue the quantization of the asymptotic phase space through the lens of \cite{Ciambelli:2024swv}. In particular, it would be rewarding to understand the connection between the central charge in the CCFT and the central charge we found in the aforementioned paper. Furthermore, we wish to explore if  the finite-distance $\beta\gamma$ CFT describing the spin-$0$ sector of the theory persists at null infinity, and the relationship between these primary fields and the asymptotic phase space data. 
\end{itemize}

In conclusion, this work bridges the gap between null physics on finite-distance hypersurfaces and asymptotic null infinity, uniting them within a cohesive framework. By drawing connections between these realms, we can leverage insights from one to address challenges in the other, with far-reaching consequences yet to be unveiled. 

\paragraph{Acknowledgements}

This work has a rich history. The idea of foliating the bulk of flat-space with null hypersurfaces to facilitate comparisons with AdS/CFT was first conceived during a discussion with Francesco Alessio at the 2019 Avogadro meeting in Naples. The development of finite-distance Carrollian tools, in collaboration with Laurent Freidel and Rob Leigh, laid the foundation for this approach. The early stages of this paper were carried out in partnership with Miguel Campiglia, to whom I owe a deep debt of gratitude. I am also thankful to Simone Speziale for joining the discussion and providing valuable encouragement and feedback. Research at Perimeter Institute is supported in part by the Government of Canada through the Department of Innovation, Science and Economic Development Canada and by the Province of Ontario through the Ministry of Colleges and Universities.

\appendix 

\section{Intrinsic Carrollian Analysis}\label{AppA}
We here review the intrinsic geometry of a null hypersurface, in the modern language of Carrollian geometry. Notation is mostly taken from  \cite{Ciambelli:2023mir}.

Intrinsically, on a finite distance null hypersurface, we can define the geometric structure given by a nowhere vanishing vector field, and a corank-1 degenerate metric
\beq
\ell^a q_{ab}=0.
\eeq
This defines the Carrollian structure \cite{Henneaux1979a}.

One can then choose a (tangent bundle) dual form to $\ell$
\beq
k_a \quad \text{such that} \quad k_a\ell^a=1.
\eeq
The ambiguity is
\beq
k_a\to k_a +j_a \quad \text{such that} \quad j_a \ell^a=0.
\eeq
This is the reason why $k_a$ is called an Ehresmann connection \cite{Ciambelli:2019lap}, it has shift symmetry.

We then construct the projector
\beq \label{defprojector}
q_a{}^b=\delta_a^b-k_a\ell^b,
\eeq
satisfying
\beq
q_a{}^b q_b{}^c=q_a{}^c\qquad \ell^a q_a{}^b=0=q_a{}^b k_b.
\eeq
The second fundamental form (also called expansion tensor or extrinsic curvature) is 
\beq
\theta_{ab}=\frac12\cL_\ell q_{ab},
\eeq
and we construct the tensor
\beq
\theta_a{}^b=\frac12 \theta q_a{}^b+\sigma_a{}^b,
\eeq
with $\sigma_a{}^a=0$. Here, $\theta$ is the expansion while $\sigma_a{}^b$ is the shear. The tensor $\theta_a{}^b$ is uniquely determined by the conditions
\beq
\theta_a{}^b k_b=0 \qquad 
\theta_a{}^b q_{bc}=\theta_{ac}.
\eeq

Then, the Raychaudhuri equation is
\beq
\left(\cL_{\ell}+\theta\right)[\theta]=\mu \theta-\sigma_a{}^b \sigma_b{}^a-R_{\ell \ell},
\eeq
with 
\beq
\mu=\kappa+\frac{\theta}{2},
\eeq
and $\kappa$ the inaffinity of $\ell$. Similarly, the Damour equation is
\beq
\quad q_a{}^b \left({\cal L}_\ell+\theta\right)\pi_b+\theta\varphi_a=(\overline{D}_b+\varphi_b)(\mu q_a{}^b-\sigma_{a}{}^b),
\eeq
with $\varphi_a=-\fL_{\ell}k_a$ the Carrollian acceleration, and $\pi_a$ the H\'aji\u{c}ek connection. In this expression, $\overline{D}_a$ is the projected horizontal derivative, that is, for a generic tensor $K_{b\cdots c}{}^{d\cdots e}$, $\overline{D}_a K_{b\cdots c}{}^{d\cdots e}=q_a{}^f q_b{}^g \cdots q_c{}^h q_i{}^d\cdots q_j{}^e D_f K_{g\cdots h}{}^{i\cdots j}$, where $D_a$ is the Carrollian connection that we define presently.

The quantities $\kappa$ and $\pi_a$ enter the intrinsic description as parts of the Carrollian connection $D_a$, which is the intrinsic connection derived from the induced Rigging connection \cite{Mars:1993mj}. This connection is extensively described in \cite{Chandrasekaran:2021hxc, Freidel:2022bai, Ciambelli:2023mir}. The latter is the torsionless connection satisfying\footnote{For completeness, although we will not need it, $D_a$ acts on $k_a$ as $(D_a+\omega_a)k_b =\bar\theta_{ab}-k_a(\pi_b+\varphi_b)$, where $\bar\theta_{ab}$ is dictated by the bulk, and $\varphi_a$ is the Carrollian acceleration defined in eq. \eqref{varphi}.}
\beqn
D_a q_{bc}= -k_b\theta_{ac}-k_c \theta_{ab}\qquad (D_a-\omega_a)\ell^b=\theta_a{}^b,
\eeqn
where $\omega_a=\kappa k_a+\pi_a$. From the ambient space perspective, this is the Rigging connection \cite{Mars:1993mj}, defined by projecting the bulk Levi-Civita connection to the hypersurface using the Rigging projector \eqref{Riggpro}. This connection does not satisfy metricity: On a Carrollian manifold, one cannot impose torsionless and metricity without constraining the underlying geometric structure,  see Appendix A of \cite{Ciambelli:2023xqk} for details.

\bibliographystyle{uiuchept}
\bibliography{LCv3.bib}

@article{Adami:2023fbm,
    author = "Adami, H. and Parvizi, A. and Sheikh-Jabbari, M. M. and Taghiloo, V. and Yavartanoo, H.",
    title = "{Hydro \& thermo dynamics at causal boundaries, examples in 3d gravity}",
    eprint = "2305.01009",
    archivePrefix = "arXiv",
    primaryClass = "hep-th",
    doi = "10.1007/JHEP07(2023)038",
    journal = "JHEP",
    volume = "07",
    pages = "038",
    year = "2023"
}

@article{Adami:2024rkr,
    author = "Adami, H. and Sheikh-Jabbari, M. M. and Taghiloo, V.",
    title = "{Gravitational stress tensor and current at null infinity in three dimensions}",
    eprint = "2405.00149",
    archivePrefix = "arXiv",
    primaryClass = "hep-th",
    doi = "10.1016/j.physletb.2024.138835",
    journal = "Phys. Lett. B",
    volume = "855",
    pages = "138835",
    year = "2024"
}

@article{Bagchi:2024gnn,
    author = "Bagchi, Arjun and Dhivakar, Prateksh and Dutta, Sudipta",
    title = "{3D Stress Tensor for Gravity in 4D Flat Spacetime}",
    eprint = "2408.05494",
    archivePrefix = "arXiv",
    primaryClass = "hep-th",
    month = "8",
    year = "2024"
}

@article{Ciambelli:2024kre,
    author = "Ciambelli, Luca and Pasterski, Sabrina and Tabor, Elisa",
    title = "{Radiation in holography}",
    eprint = "2404.02146",
    archivePrefix = "arXiv",
    primaryClass = "hep-th",
    doi = "10.1007/JHEP09(2024)124",
    journal = "JHEP",
    volume = "09",
    pages = "124",
    year = "2024"
}

@article{Cachazo:2014fwa,
    author = "Cachazo, Freddy and Strominger, Andrew",
    title = "{Evidence for a New Soft Graviton Theorem}",
    eprint = "1404.4091",
    archivePrefix = "arXiv",
    primaryClass = "hep-th",
    month = "4",
    year = "2014"
}

@article{Hartman:2016lgu,
    author = "Hartman, Thomas and Kundu, Sandipan and Tajdini, Amirhossein",
    title = "{Averaged Null Energy Condition from Causality}",
    eprint = "1610.05308",
    archivePrefix = "arXiv",
    primaryClass = "hep-th",
    doi = "10.1007/JHEP07(2017)066",
    journal = "JHEP",
    volume = "07",
    pages = "066",
    year = "2017"
}

@article{Ceyhan:2018zfg,
    author = "Ceyhan, Fikret and Faulkner, Thomas",
    title = "{Recovering the QNEC from the ANEC}",
    eprint = "1812.04683",
    archivePrefix = "arXiv",
    primaryClass = "hep-th",
    doi = "10.1007/s00220-020-03751-y",
    journal = "Commun. Math. Phys.",
    volume = "377",
    number = "2",
    pages = "999--1045",
    year = "2020"
}

@article{Balakrishnan:2017bjg,
    author = "Balakrishnan, Srivatsan and Faulkner, Thomas and Khandker, Zuhair U. and Wang, Huajia",
    title = "{A General Proof of the Quantum Null Energy Condition}",
    eprint = "1706.09432",
    archivePrefix = "arXiv",
    primaryClass = "hep-th",
    doi = "10.1007/JHEP09(2019)020",
    journal = "JHEP",
    volume = "09",
    pages = "020",
    year = "2019"
}

@article{Kontou:2020bta,
    author = "Kontou, Eleni-Alexandra and Sanders, Ko",
    title = "{Energy conditions in general relativity and quantum field theory}",
    eprint = "2003.01815",
    archivePrefix = "arXiv",
    primaryClass = "gr-qc",
    doi = "10.1088/1361-6382/ab8fcf",
    journal = "Class. Quant. Grav.",
    volume = "37",
    number = "19",
    pages = "193001",
    year = "2020"
}

@article{Ciambelli:2023xqk,
    author = "Ciambelli, Luca",
    title = "{Dynamics of Carrollian scalar fields}",
    eprint = "2311.04113",
    archivePrefix = "arXiv",
    primaryClass = "hep-th",
    doi = "10.1088/1361-6382/ad5bb5",
    journal = "Class. Quant. Grav.",
    volume = "41",
    number = "16",
    pages = "165011",
    year = "2024"
}

@article{Donnay:2022aba,
    author = "Donnay, Laura and Fiorucci, Adrien and Herfray, Yannick and Ruzziconi, Romain",
    title = "{Carrollian Perspective on Celestial Holography}",
    eprint = "2202.04702",
    archivePrefix = "arXiv",
    primaryClass = "hep-th",
    doi = "10.1103/PhysRevLett.129.071602",
    journal = "Phys. Rev. Lett.",
    volume = "129",
    number = "7",
    pages = "071602",
    year = "2022"
}

@article{Redondo-Yuste:2022czg,
    author = "Redondo-Yuste, Jaime and Lehner, Luis",
    title = "{Non-linear black hole dynamics and Carrollian fluids}",
    eprint = "2212.06175",
    archivePrefix = "arXiv",
    primaryClass = "gr-qc",
    doi = "10.1007/JHEP02(2023)240",
    journal = "JHEP",
    volume = "02",
    pages = "240",
    year = "2023"
}

@article{Lewkowycz:2013nqa,
    author = "Lewkowycz, Aitor and Maldacena, Juan",
    title = "{Generalized gravitational entropy}",
    eprint = "1304.4926",
    archivePrefix = "arXiv",
    primaryClass = "hep-th",
    doi = "10.1007/JHEP08(2013)090",
    journal = "JHEP",
    volume = "08",
    pages = "090",
    year = "2013"
}

@article{Li:2010dr,
    author = "Li, Wei and Takayanagi, Tadashi",
    title = "{Holography and Entanglement in Flat Spacetime}",
    eprint = "1010.3700",
    archivePrefix = "arXiv",
    primaryClass = "hep-th",
    reportNumber = "IPMU10-0184",
    doi = "10.1103/PhysRevLett.106.141301",
    journal = "Phys. Rev. Lett.",
    volume = "106",
    pages = "141301",
    year = "2011"
}

@article{Apolo:2020bld,
    author = "Apolo, Luis and Jiang, Hongliang and Song, Wei and Zhong, Yuan",
    title = "{Swing surfaces and holographic entanglement beyond AdS/CFT}",
    eprint = "2006.10740",
    archivePrefix = "arXiv",
    primaryClass = "hep-th",
    doi = "10.1007/JHEP12(2020)064",
    journal = "JHEP",
    volume = "12",
    pages = "064",
    year = "2020"
}

@article{Dong:2016eik,
    author = "Dong, Xi and Harlow, Daniel and Wall, Aron C.",
    title = "{Reconstruction of Bulk Operators within the Entanglement Wedge in Gauge-Gravity Duality}",
    eprint = "1601.05416",
    archivePrefix = "arXiv",
    primaryClass = "hep-th",
    reportNumber = "NSF-KITP-16-005",
    doi = "10.1103/PhysRevLett.117.021601",
    journal = "Phys. Rev. Lett.",
    volume = "117",
    number = "2",
    pages = "021601",
    year = "2016"
}

@article{VanRaamsdonk:2010pw,
    author = "Van Raamsdonk, Mark",
    title = "{Building up spacetime with quantum entanglement}",
    eprint = "1005.3035",
    archivePrefix = "arXiv",
    primaryClass = "hep-th",
    doi = "10.1142/S0218271810018529",
    journal = "Gen. Rel. Grav.",
    volume = "42",
    pages = "2323--2329",
    year = "2010"
}

@article{Ryu:2006bv,
    author = "Ryu, Shinsei and Takayanagi, Tadashi",
    title = "{Holographic derivation of entanglement entropy from AdS/CFT}",
    eprint = "hep-th/0603001",
    archivePrefix = "arXiv",
    reportNumber = "NSF-KITP-06-11",
    doi = "10.1103/PhysRevLett.96.181602",
    journal = "Phys. Rev. Lett.",
    volume = "96",
    pages = "181602",
    year = "2006"
}

@article{Rignon-Bret:2024zhj,
    author = "Rignon-Bret, Antoine",
    title = "{Black hole thermodynamic potentials for asymptotic observers}",
    eprint = "2406.15843",
    archivePrefix = "arXiv",
    primaryClass = "gr-qc",
    doi = "10.1103/PhysRevD.110.124002",
    journal = "Phys. Rev. D",
    volume = "110",
    number = "12",
    pages = "124002",
    year = "2024"
}

@article{He:2014laa,
    author = "He, Temple and Lysov, Vyacheslav and Mitra, Prahar and Strominger, Andrew",
    title = "{BMS supertranslations and Weinberg\textquoteright{}s soft graviton theorem}",
    eprint = "1401.7026",
    archivePrefix = "arXiv",
    primaryClass = "hep-th",
    doi = "10.1007/JHEP05(2015)151",
    journal = "JHEP",
    volume = "05",
    pages = "151",
    year = "2015"
}

@article{Campiglia:2014yka,
    author = "Campiglia, Miguel and Laddha, Alok",
    title = "{Asymptotic symmetries and subleading soft graviton theorem}",
    eprint = "1408.2228",
    archivePrefix = "arXiv",
    primaryClass = "hep-th",
    doi = "10.1103/PhysRevD.90.124028",
    journal = "Phys. Rev. D",
    volume = "90",
    number = "12",
    pages = "124028",
    year = "2014"
}

@article{Bhattacharyya:2007vjd,
    author = "Bhattacharyya, Sayantani and Hubeny, Veronika E and Minwalla, Shiraz and Rangamani, Mukund",
    title = "{Nonlinear Fluid Dynamics from Gravity}",
    eprint = "0712.2456",
    archivePrefix = "arXiv",
    primaryClass = "hep-th",
    reportNumber = "TIFR-TH-07-44, DCPT-07-73, NI07097",
    doi = "10.1088/1126-6708/2008/02/045",
    journal = "JHEP",
    volume = "02",
    pages = "045",
    year = "2008"
}

@article{Pasterski:2023ikd,
    author = "Pasterski, Sabrina",
    title = "{A Chapter on Celestial Holography}",
    eprint = "2310.04932",
    archivePrefix = "arXiv",
    primaryClass = "hep-th",
    month = "10",
    year = "2023"
}

@article{Witten:1998qj,
    author = "Witten, Edward",
    title = "{Anti-de Sitter space and holography}",
    eprint = "hep-th/9802150",
    archivePrefix = "arXiv",
    reportNumber = "IASSNS-HEP-98-15",
    doi = "10.4310/ATMP.1998.v2.n2.a2",
    journal = "Adv. Theor. Math. Phys.",
    volume = "2",
    pages = "253--291",
    year = "1998"
}

@article{Riello:2024uvs,
    author = "Riello, Aldo and Freidel, Laurent",
    title = "{Renormalization of conformal infinity as a stretched horizon}",
    eprint = "2402.03097",
    archivePrefix = "arXiv",
    primaryClass = "gr-qc",
    doi = "10.1088/1361-6382/ad5cbb",
    journal = "Class. Quant. Grav.",
    volume = "41",
    number = "17",
    pages = "175013",
    year = "2024"
}

@article{Geiller:2022vto,
    author = "Geiller, Marc and Zwikel, C\'eline",
    title = "{The partial Bondi gauge: Further enlarging the asymptotic structure of gravity}",
    eprint = "2205.11401",
    archivePrefix = "arXiv",
    primaryClass = "hep-th",
    doi = "10.21468/SciPostPhys.13.5.108",
    journal = "SciPost Phys.",
    volume = "13",
    pages = "108",
    year = "2022"
}

@article{Geiller:2024amx,
    author = "Geiller, Marc and Zwikel, C\'eline",
    title = "{The partial Bondi gauge: Gauge fixings and asymptotic charges}",
    eprint = "2401.09540",
    archivePrefix = "arXiv",
    primaryClass = "hep-th",
    doi = "10.21468/SciPostPhys.16.3.076",
    journal = "SciPost Phys.",
    volume = "16",
    pages = "076",
    year = "2024"
}

@article{Kraus:2024gso,
    author = "Kraus, Per and Myers, Richard M.",
    title = "{Carrollian Partition Functions and the Flat Limit of AdS}",
    eprint = "2407.13668",
    archivePrefix = "arXiv",
    primaryClass = "hep-th",
    month = "7",
    year = "2024"
}

@article{Bagchi:2015wna,
    author = "Bagchi, Arjun and Grumiller, Daniel and Merbis, Wout",
    title = "{Stress tensor correlators in three-dimensional gravity}",
    eprint = "1507.05620",
    archivePrefix = "arXiv",
    primaryClass = "hep-th",
    reportNumber = "MIT-CTP-4694, TUW-15-13",
    doi = "10.1103/PhysRevD.93.061502",
    journal = "Phys. Rev. D",
    volume = "93",
    number = "6",
    pages = "061502",
    year = "2016"
}

@article{Ruzziconi:2024kzo,
    author = "Ruzziconi, Romain and Saha, Amartya",
    title = "{Holographic Carrollian Currents for Massless Scattering}",
    eprint = "2411.04902",
    archivePrefix = "arXiv",
    primaryClass = "hep-th",
    month = "11",
    year = "2024"
}

@article{Bhambure:2024ftz,
    author = "Bhambure, Jay and Krishna, Hare",
    title = "{A stress tensor for asymptotically flat spacetime}",
    eprint = "2412.08588",
    archivePrefix = "arXiv",
    primaryClass = "hep-th",
    reportNumber = "YITP-SB-2024-33",
    month = "12",
    year = "2024"
}

@article{Kapec:2017gsg,
    author = "Kapec, Daniel and Mitra, Prahar",
    title = "{A $d$-Dimensional Stress Tensor for Mink$_{d+2}$ Gravity}",
    eprint = "1711.04371",
    archivePrefix = "arXiv",
    primaryClass = "hep-th",
    doi = "10.1007/JHEP05(2018)186",
    journal = "JHEP",
    volume = "05",
    pages = "186",
    year = "2018"
}

@article{Kapec:2016jld,
    author = "Kapec, Daniel and Mitra, Prahar and Raclariu, Ana-Maria and Strominger, Andrew",
    title = "{2D Stress Tensor for 4D Gravity}",
    eprint = "1609.00282",
    archivePrefix = "arXiv",
    primaryClass = "hep-th",
    doi = "10.1103/PhysRevLett.119.121601",
    journal = "Phys. Rev. Lett.",
    volume = "119",
    number = "12",
    pages = "121601",
    year = "2017"
}

@article{deHaro:2000vlm,
    author = "de Haro, Sebastian and Solodukhin, Sergey N. and Skenderis, Kostas",
    title = "{Holographic reconstruction of space-time and renormalization in the AdS / CFT correspondence}",
    eprint = "hep-th/0002230",
    archivePrefix = "arXiv",
    reportNumber = "SPIN-2000-05, ITP-UU-00-03, PUTP-1921",
    doi = "10.1007/s002200100381",
    journal = "Commun. Math. Phys.",
    volume = "217",
    pages = "595--622",
    year = "2001"
}

@article{Emparan:1999pm,
    author = "Emparan, Roberto and Johnson, Clifford V. and Myers, Robert C.",
    title = "{Surface terms as counterterms in the AdS / CFT correspondence}",
    eprint = "hep-th/9903238",
    archivePrefix = "arXiv",
    reportNumber = "DTP-99-21, UK-99-04, MCGILL-99-12, EHU-FT-9906",
    doi = "10.1103/PhysRevD.60.104001",
    journal = "Phys. Rev. D",
    volume = "60",
    pages = "104001",
    year = "1999"
}

@article{Balasubramanian:1999re,
    author = "Balasubramanian, Vijay and Kraus, Per",
    title = "{A Stress tensor for Anti-de Sitter gravity}",
    eprint = "hep-th/9902121",
    archivePrefix = "arXiv",
    reportNumber = "HUTP-99-A002, EFI-99-6, NSF-ITP-98-132",
    doi = "10.1007/s002200050764",
    journal = "Commun. Math. Phys.",
    volume = "208",
    pages = "413--428",
    year = "1999"
}

@article{Ashtekar:2024stm,
    author = "Ashtekar, Abhay and Speziale, Simone",
    title = "{Null infinity and horizons: A new approach to fluxes and charges}",
    eprint = "2407.03254",
    archivePrefix = "arXiv",
    primaryClass = "hep-th",
    doi = "10.1103/PhysRevD.110.044049",
    journal = "Phys. Rev. D",
    volume = "110",
    number = "4",
    pages = "044049",
    year = "2024"
}

@article{Compere:2019bua,
    author = "Comp\`ere, Geoffrey and Fiorucci, Adrien and Ruzziconi, Romain",
    title = "{The $\Lambda$-BMS$_4$ group of dS$_4$ and new boundary conditions for AdS$_4$}",
    eprint = "1905.00971",
    archivePrefix = "arXiv",
    primaryClass = "gr-qc",
    doi = "10.1088/1361-6382/ab3d4b",
    journal = "Class. Quant. Grav.",
    volume = "36",
    number = "19",
    pages = "195017",
    year = "2019",
    note = "[Erratum: Class.Quant.Grav. 38, 229501 (2021)]"
}

@article{Campiglia:2024uqq,
    author = "Campiglia, Miguel and Sudhakar, Adarsh",
    title = "{Gravitational Poisson brackets at null infinity compatible with smooth superrotations}",
    eprint = "2408.13067",
    archivePrefix = "arXiv",
    primaryClass = "gr-qc",
    doi = "10.1007/JHEP12(2024)170",
    journal = "JHEP",
    volume = "12",
    pages = "170",
    year = "2024"
}

@article{He:2024vlp,
    author = "He, Temple and Raclariu, Ana-Maria and Zurek, Kathryn M.",
    title = "{An Infrared On-Shell Action and its Implications for Soft Charge Fluctuations in Asymptotically Flat Spacetimes}",
    eprint = "2408.01485",
    archivePrefix = "arXiv",
    primaryClass = "hep-th",
    reportNumber = "CALT-TH 2024-028",
    month = "8",
    year = "2024"
}

@article{Ciambelli:2024swv,
    author = "Ciambelli, Luca and Freidel, Laurent and Leigh, Robert G.",
    title = "{Quantum null geometry and gravity}",
    eprint = "2407.11132",
    archivePrefix = "arXiv",
    primaryClass = "hep-th",
    doi = "10.1007/JHEP12(2024)028",
    journal = "JHEP",
    volume = "12",
    pages = "028",
    year = "2024"
}

@article{Kapec:2016aqd,
    author = "Kapec, Daniel and Raclariu, Ana-Maria and Strominger, Andrew",
    title = "{Area, Entanglement Entropy and Supertranslations at Null Infinity}",
    eprint = "1603.07706",
    archivePrefix = "arXiv",
    primaryClass = "hep-th",
    doi = "10.1088/1361-6382/aa7f12",
    journal = "Class. Quant. Grav.",
    volume = "34",
    number = "16",
    pages = "165007",
    year = "2017"
}

@article{Hartong:2015xda,
    author = "Hartong, Jelle",
    title = "{Gauging the Carroll Algebra and Ultra-Relativistic Gravity}",
    eprint = "1505.05011",
    archivePrefix = "arXiv",
    primaryClass = "hep-th",
    doi = "10.1007/JHEP08(2015)069",
    journal = "JHEP",
    volume = "08",
    pages = "069",
    year = "2015"
}

@article{Compere:2018ylh,
    author = "Comp\`ere, Geoffrey and Fiorucci, Adrien and Ruzziconi, Romain",
    title = "{Superboost transitions, refraction memory and super-Lorentz charge algebra}",
    eprint = "1810.00377",
    archivePrefix = "arXiv",
    primaryClass = "hep-th",
    doi = "10.1007/JHEP11(2018)200",
    journal = "JHEP",
    volume = "11",
    pages = "200",
    year = "2018",
    note = "[Erratum: JHEP 04, 172 (2020)]"
}

@article{Ciambelli:2023mvj,
    author = "Ciambelli, Luca and Lehner, Luis",
    title = "{Fluid-gravity correspondence and causal first-order relativistic viscous hydrodynamics}",
    eprint = "2310.15427",
    archivePrefix = "arXiv",
    primaryClass = "hep-th",
    doi = "10.1103/PhysRevD.108.126019",
    journal = "Phys. Rev. D",
    volume = "108",
    number = "12",
    pages = "126019",
    year = "2023"
}

@article{Campiglia:2020qvc,
    author = "Campiglia, Miguel and Peraza, Javier",
    title = "{Generalized BMS charge algebra}",
    eprint = "2002.06691",
    archivePrefix = "arXiv",
    primaryClass = "gr-qc",
    doi = "10.1103/PhysRevD.101.104039",
    journal = "Phys. Rev. D",
    volume = "101",
    number = "10",
    pages = "104039",
    year = "2020"
}

@article{Barnich:2010eb,
    author = "Barnich, Glenn and Troessaert, Cedric",
    title = "{Aspects of the BMS/CFT correspondence}",
    eprint = "1001.1541",
    archivePrefix = "arXiv",
    primaryClass = "hep-th",
    reportNumber = "ULB-TH-09-28",
    doi = "10.1007/JHEP05(2010)062",
    journal = "JHEP",
    volume = "05",
    pages = "062",
    year = "2010"
}

@article{Donnay:2019jiz,
    author = "Donnay, Laura and Marteau, Charles",
    title = "{Carrollian Physics at the Black Hole Horizon}",
    eprint = "1903.09654",
    archivePrefix = "arXiv",
    primaryClass = "hep-th",
    doi = "10.1088/1361-6382/ab2fd5",
    journal = "Class. Quant. Grav.",
    volume = "36",
    number = "16",
    pages = "165002",
    year = "2019"
}

@article{Ciambelli:2018wre,
    author = "Ciambelli, Luca and Marteau, Charles and Petkou, Anastasios C. and Petropoulos, P. Marios and Siampos, Konstantinos",
    title = "{Flat holography and Carrollian fluids}",
    eprint = "1802.06809",
    archivePrefix = "arXiv",
    primaryClass = "hep-th",
    reportNumber = "CPHT-RR049.082017, CERN-TH-2017-229",
    doi = "10.1007/JHEP07(2018)165",
    journal = "JHEP",
    volume = "07",
    pages = "165",
    year = "2018"
}

@article{Hopfmuller:2018fni,
    author = {Hopfm\"uller, Florian and Freidel, Laurent},
    title = "{Null Conservation Laws for Gravity}",
    eprint = "1802.06135",
    archivePrefix = "arXiv",
    primaryClass = "gr-qc",
    doi = "10.1103/PhysRevD.97.124029",
    journal = "Phys. Rev. D",
    volume = "97",
    number = "12",
    pages = "124029",
    year = "2018"
}

@article{Wall:2011hj,
    author = "Wall, Aron C.",
    title = "{A proof of the generalized second law for rapidly changing fields and arbitrary horizon slices}",
    eprint = "1105.3445",
    archivePrefix = "arXiv",
    primaryClass = "gr-qc",
    doi = "10.1103/PhysRevD.85.104049",
    journal = "Phys. Rev. D",
    volume = "85",
    pages = "104049",
    year = "2012",
    note = "[Erratum: Phys.Rev.D 87, 069904 (2013)]"
}

@article{Gourgoulhon:2005ng,
    author = "Gourgoulhon, Eric and Jaramillo, Jose Luis",
    title = "{A 3+1 perspective on null hypersurfaces and isolated horizons}",
    eprint = "gr-qc/0503113",
    archivePrefix = "arXiv",
    doi = "10.1016/j.physrep.2005.10.005",
    journal = "Phys. Rept.",
    volume = "423",
    pages = "159--294",
    year = "2006"
}

@article{Wieland:2017zkf,
    author = "Wieland, Wolfgang",
    title = "{New boundary variables for classical and quantum gravity on a null surface}",
    eprint = "1704.07391",
    archivePrefix = "arXiv",
    primaryClass = "gr-qc",
    doi = "10.1088/1361-6382/aa8d06",
    journal = "Class. Quant. Grav.",
    volume = "34",
    number = "21",
    pages = "215008",
    year = "2017"
}

@article{Chandrasekaran:2021hxc,
    author = "Chandrasekaran, Venkatesa and Flanagan, Eanna E. and Shehzad, Ibrahim and Speranza, Antony J.",
    title = "{Brown-York charges at null boundaries}",
    eprint = "2109.11567",
    archivePrefix = "arXiv",
    primaryClass = "hep-th",
    doi = "10.1007/JHEP01(2022)029",
    journal = "JHEP",
    volume = "01",
    pages = "029",
    year = "2022"
}

@article{Hopfmuller:2016scf,
    author = {Hopfm\"uller, Florian and Freidel, Laurent},
    title = "{Gravity Degrees of Freedom on a Null Surface}",
    eprint = "1611.03096",
    archivePrefix = "arXiv",
    primaryClass = "gr-qc",
    doi = "10.1103/PhysRevD.95.104006",
    journal = "Phys. Rev. D",
    volume = "95",
    number = "10",
    pages = "104006",
    year = "2017"
}

@article{Chandrasekaran:2018aop,
    author = "Chandrasekaran, Venkatesa and Flanagan, \'Eanna \'E. and Prabhu, Kartik",
    title = "{Symmetries and charges of general relativity at null boundaries}",
    eprint = "1807.11499",
    archivePrefix = "arXiv",
    primaryClass = "hep-th",
    doi = "10.1007/JHEP11(2018)125",
    journal = "JHEP",
    volume = "11",
    pages = "125",
    year = "2018"
}

@article{Lehner:2016vdi,
    author = "Lehner, Luis and Myers, Robert C. and Poisson, Eric and Sorkin, Rafael D.",
    title = "{Gravitational action with null boundaries}",
    eprint = "1609.00207",
    archivePrefix = "arXiv",
    primaryClass = "hep-th",
    doi = "10.1103/PhysRevD.94.084046",
    journal = "Phys. Rev. D",
    volume = "94",
    number = "8",
    pages = "084046",
    year = "2016"
}

@article{Ciambelli:2019lap,
    author = "Ciambelli, Luca and Leigh, Robert G. and Marteau, Charles and Petropoulos, P. Marios",
    title = "{Carroll Structures, Null Geometry and Conformal Isometries}",
    eprint = "1905.02221",
    archivePrefix = "arXiv",
    primaryClass = "hep-th",
    reportNumber = "CPHT-RR025.052019, CPHT-RR010.022019",
    doi = "10.1103/PhysRevD.100.046010",
    journal = "Phys. Rev. D",
    volume = "100",
    number = "4",
    pages = "046010",
    year = "2019"
}

@article{Mars:1993mj,
    author = "Mars, Marc and Senovilla, Jose M. M.",
    title = "{Geometry of general hypersurfaces in space-time: Junction conditions}",
    eprint = "gr-qc/0201054",
    archivePrefix = "arXiv",
    doi = "10.1088/0264-9381/10/9/026",
    journal = "Class. Quant. Grav.",
    volume = "10",
    pages = "1865--1897",
    year = "1993"
}

@article{Chandrasekaran:2020wwn,
    author = "Chandrasekaran, Venkatesa and Speranza, Antony J.",
    title = "{Anomalies in gravitational charge algebras of null boundaries and black hole entropy}",
    eprint = "2009.10739",
    archivePrefix = "arXiv",
    primaryClass = "hep-th",
    doi = "10.1007/JHEP01(2021)137",
    journal = "JHEP",
    volume = "01",
    pages = "137",
    year = "2021"
}

@article{Adami:2021kvx,
    author = "Adami, H. and Sheikh-Jabbari, M. M. and Taghiloo, V. and Yavartanoo, H.",
    title = "{Null surface thermodynamics}",
    eprint = "2110.04224",
    archivePrefix = "arXiv",
    primaryClass = "hep-th",
    doi = "10.1103/PhysRevD.105.066004",
    journal = "Phys. Rev. D",
    volume = "105",
    number = "6",
    pages = "066004",
    year = "2022"
}

@article{Duval:2014uva,
    author = "Duval, C. and Gibbons, G. W. and Horvathy, P. A.",
    title = "{Conformal Carroll groups and BMS symmetry}",
    eprint = "1402.5894",
    archivePrefix = "arXiv",
    primaryClass = "gr-qc",
    doi = "10.1088/0264-9381/31/9/092001",
    journal = "Class. Quant. Grav.",
    volume = "31",
    pages = "092001",
    year = "2014"
}

@article{Donnay:2016ejv,
    author = "Donnay, Laura and Giribet, Gaston and Gonz\'alez, Hern\'an A. and Pino, Miguel",
    title = "{Extended Symmetries at the Black Hole Horizon}",
    eprint = "1607.05703",
    archivePrefix = "arXiv",
    primaryClass = "hep-th",
    doi = "10.1007/JHEP09(2016)100",
    journal = "JHEP",
    volume = "09",
    pages = "100",
    year = "2016"
}

@article{Odak:2023pga,
    author = "Odak, Gloria and Rignon-Bret, Antoine and Speziale, Simone",
    title = "{General gravitational charges on null hypersurfaces}",
    eprint = "2309.03854",
    archivePrefix = "arXiv",
    primaryClass = "gr-qc",
    month = "9",
    year = "2023"
}

@article{Chandrasekaran:2023vzb,
    author = "Chandrasekaran, Venkatesa and Flanagan, Eanna E.",
    title = "{The gravitational phase space of horizons in general relativity}",
    eprint = "2309.03871",
    archivePrefix = "arXiv",
    primaryClass = "gr-qc",
    month = "9",
    year = "2023"
}

@article{Ciambelli:2023mir,
    author = "Ciambelli, Luca and Freidel, Laurent and Leigh, Robert G.",
    title = "{Null Raychaudhuri: Canonical Structure and the Dressing Time}",
    eprint = "2309.03932",
    archivePrefix = "arXiv",
    primaryClass = "hep-th",
    month = "9",
    year = "2023"
}

@article{Reisenberger:2007ku,
    author = "Reisenberger, Michael P.",
    title = "{The Poisson bracket on free null initial data for gravity}",
    eprint = "0712.2541",
    archivePrefix = "arXiv",
    primaryClass = "gr-qc",
    doi = "10.1103/PhysRevLett.101.211101",
    journal = "Phys. Rev. Lett.",
    volume = "101",
    pages = "211101",
    year = "2008"
}

@article{Reisenberger:2007pq,
    author = "Reisenberger, Michael P.",
    title = "{The Symplectic 2-form and Poisson bracket of null canonical gravity}",
    eprint = "gr-qc/0703134",
    archivePrefix = "arXiv",
    month = "3",
    year = "2007"
}

@article{Adami:2021nnf,
    author = "Adami, H. and Grumiller, D. and Sheikh-Jabbari, M. M. and Taghiloo, V. and Yavartanoo, H. and Zwikel, C.",
    title = "{Null boundary phase space: slicings, news \& memory}",
    eprint = "2110.04218",
    archivePrefix = "arXiv",
    primaryClass = "hep-th",
    doi = "10.1007/JHEP11(2021)155",
    journal = "JHEP",
    volume = "11",
    pages = "155",
    year = "2021"
}

@article{Freidel:2022bai,
    author = "Freidel, Laurent and Jai-akson, Puttarak",
    title = "{Carrollian hydrodynamics from symmetries}",
    eprint = "2209.03328",
    archivePrefix = "arXiv",
    primaryClass = "hep-th",
    doi = "10.1088/1361-6382/acb194",
    journal = "Class. Quant. Grav.",
    volume = "40",
    number = "5",
    pages = "055009",
    year = "2023"
}

@article{Chandrasekaran:2021vyu,
    author = "Chandrasekaran, Venkatesa and Flanagan, Eanna E. and Shehzad, Ibrahim and Speranza, Antony J.",
    title = "{A general framework for gravitational charges and holographic renormalization}",
    eprint = "2111.11974",
    archivePrefix = "arXiv",
    primaryClass = "gr-qc",
    doi = "10.1142/S0217751X22501056",
    journal = "Int. J. Mod. Phys. A",
    volume = "37",
    number = "17",
    pages = "2250105",
    year = "2022"
}

@article{Bondi,
author = {Bondi, Hermann  and Van der Burg, M. G. J.  and Metzner, A. W. K. },
title = {Gravitational waves in general relativity, VII. Waves from axi-symmetric isolated system},
journal = {Proceedings of the Royal Society of London. Series A. Mathematical and Physical Sciences},
volume = {269},
number = {1336},
pages = {21-52},
year = {1962},
doi = {10.1098/rspa.1962.0161},
}

@book{Strominger:2017zoo,
    author = "Strominger, Andrew",
    title = "{Lectures on the Infrared Structure of Gravity and Gauge Theory}",
    eprint = "1703.05448",
    archivePrefix = "arXiv",
    primaryClass = "hep-th",
    month = "3",
    year = "2017",
    publisher = "Princeton University Press",
}

@article{Pasterski:2021rjz,
    author = "Pasterski, Sabrina",
    title = "{Lectures on celestial amplitudes}",
    eprint = "2108.04801",
    archivePrefix = "arXiv",
    primaryClass = "hep-th",
    doi = "10.1140/epjc/s10052-021-09846-7",
    journal = "Eur. Phys. J. C",
    volume = "81",
    number = "12",
    pages = "1062",
    year = "2021"
}

@article{Raclariu:2021zjz,
    author = "Raclariu, Ana-Maria",
    title = "{Lectures on Celestial Holography}",
    eprint = "2107.02075",
    archivePrefix = "arXiv",
    primaryClass = "hep-th",
    month = "7",
    year = "2021"
}

@article{Price:1986yy,
    author = "Price, R. H. and Thorne, K. S.",
    title = "{Membrane Viewpoint on Black Holes: Properties and Evolution of the Stretched Horizon}",
    doi = "10.1103/PhysRevD.33.915",
    journal = "Phys. Rev. D",
    volume = "33",
    pages = "915--941",
    year = "1986"
}

@article{Sachs:1961zz,
    author = "Sachs, R. K.",
    title = "{Gravitational waves in general relativity. 6. The outgoing radiation condition}",
    doi = "10.1098/rspa.1961.0202",
    journal = "Proc. Roy. Soc. Lond. A",
    volume = "264",
    pages = "309--338",
    year = "1961"
}

@article{Bousso:2015mna,
    author = "Bousso, Raphael and Fisher, Zachary and Leichenauer, Stefan and Wall, Aron C.",
    title = "{Quantum focusing conjecture}",
    eprint = "1506.02669",
    archivePrefix = "arXiv",
    primaryClass = "hep-th",
    doi = "10.1103/PhysRevD.93.064044",
    journal = "Phys. Rev. D",
    volume = "93",
    number = "6",
    pages = "064044",
    year = "2016"
}

@article{Sheikh-Jabbari:2022mqi,
    author = "Sheikh-Jabbari, M. M.",
    title = "{On symplectic form for null boundary phase space}",
    eprint = "2209.05043",
    archivePrefix = "arXiv",
    primaryClass = "gr-qc",
    doi = "10.1007/s10714-022-02997-2",
    journal = "Gen. Rel. Grav.",
    volume = "54",
    number = "11",
    pages = "140",
    year = "2022"
}

@article{Saha:2023abr,
    author = "Saha, Amartya",
    title = "{w$_{1+\infty}$ and Carrollian holography}",
    eprint = "2308.03673",
    archivePrefix = "arXiv",
    primaryClass = "hep-th",
    doi = "10.1007/JHEP05(2024)145",
    journal = "JHEP",
    volume = "05",
    pages = "145",
    year = "2024"
}

@article{Saha:2023hsl,
    author = "Saha, Amartya",
    title = "{Carrollian approach to 1 + 3D flat holography}",
    eprint = "2304.02696",
    archivePrefix = "arXiv",
    primaryClass = "hep-th",
    doi = "10.1007/JHEP06(2023)051",
    journal = "JHEP",
    volume = "06",
    pages = "051",
    year = "2023"
}

@article{Adami:2020ugu,
    author = "Adami, H. and Sheikh-Jabbari, M. M. and Taghiloo, V. and Yavartanoo, H. and Zwikel, C.",
    title = "{Symmetries at null boundaries: two and three dimensional gravity cases}",
    eprint = "2007.12759",
    archivePrefix = "arXiv",
    primaryClass = "hep-th",
    doi = "10.1007/JHEP10(2020)107",
    journal = "JHEP",
    volume = "10",
    pages = "107",
    year = "2020"
}

@article{Casini:2017roe,
    author = "Casini, Horacio and Teste, Eduardo and Torroba, Gonzalo",
    title = "{Modular Hamiltonians on the null plane and the Markov property of the vacuum state}",
    eprint = "1703.10656",
    archivePrefix = "arXiv",
    primaryClass = "hep-th",
    doi = "10.1088/1751-8121/aa7eaa",
    journal = "J. Phys. A",
    volume = "50",
    number = "36",
    pages = "364001",
    year = "2017"
}

@article{Bousso:1999xy,
    author = "Bousso, Raphael",
    title = "{A Covariant entropy conjecture}",
    eprint = "hep-th/9905177",
    archivePrefix = "arXiv",
    reportNumber = "SU-ITP-99-23",
    doi = "10.1088/1126-6708/1999/07/004",
    journal = "JHEP",
    volume = "07",
    pages = "004",
    year = "1999"
}

@article{Freidel:2021fxf,
    author = "Freidel, Laurent and Oliveri, Roberto and Pranzetti, Daniele and Speziale, Simone",
    title = "{The Weyl BMS group and Einstein\textquoteright{}s equations}",
    eprint = "2104.05793",
    archivePrefix = "arXiv",
    primaryClass = "hep-th",
    doi = "10.1007/JHEP07(2021)170",
    journal = "JHEP",
    volume = "07",
    pages = "170",
    year = "2021"
}

@article{Ashtekar1981,
	author = {Ashtekar, A. and Streubel, M.},
	doi = {10.1098/rspa.1981.0109},
	journal = {Proc. Roy. Soc. Lond. A},
	pages = {585--607},
	title = {{Symplectic Geometry of Radiative Modes and Conserved Quantities at Null Infinity}},
	volume = {376},
	year = {1981}}

@article{Damour1979,
	author = {T. Damour},
	journal = {Th\'ese de Doctorat d'Etat, Universit\'e Pierre et Marie Curie, Paris VI},
	title = {{Quelques propri\'et\'es m\'ecaniques, \'electromagn\'etiques, thermodynamiques et quantiques des trous noirs}},
	year = {1979}}

@article{Ciambelli:2018ojf,
	archiveprefix = {arXiv},
	author = {Ciambelli, Luca and Marteau, Charles},
	doi = {10.1088/1361-6382/ab0d37},
	eprint = {1810.11037},
	journal = {Class. Quant. Grav.},
	number = {8},
	pages = {085004},
	primaryclass = {hep-th},
	reportnumber = {CPHT-RR101.102018},
	title = {{Carrollian conservation laws and Ricci-flat gravity}},
	volume = {36},
	year = {2019}}

@article{Henneaux1979a,
	author = {Henneaux, M.},
	journal = {Bull.Soc.Math.Belg. 31 47-63},
	title = {{Geometry of zero signature spacetime}},
	year = {1979}}

@article{Bousso:2017xyo,
    author = "Bousso, Raphael and Chandrasekaran, Venkatesa and Halpern, Illan F. and Wall, Aron",
    title = "{Asymptotic Charges Cannot Be Measured in Finite Time}",
    eprint = "1709.08632",
    archivePrefix = "arXiv",
    primaryClass = "hep-th",
    doi = "10.1103/PhysRevD.97.046014",
    journal = "Phys. Rev. D",
    volume = "97",
    number = "4",
    pages = "046014",
    year = "2018"
}

@article{Hayward:1993my,
    author = "Hayward, G.",
    title = "{Gravitational action for space-times with nonsmooth boundaries}",
    doi = "10.1103/PhysRevD.47.3275",
    journal = "Phys. Rev. D",
    volume = "47",
    pages = "3275--3280",
    year = "1993"
}

@article{brown1993quasilocal,
  title={Quasilocal energy and conserved charges derived from the gravitational action},
  author={Brown, J David and York Jr, James W},
  journal={Physical Review D},
  volume={47},
  number={4},
  pages={1407},
  year={1993},
  publisher={APS}
}

@article{Campoleoni:2018ltl,
    author = "Campoleoni, Andrea and Ciambelli, Luca and Marteau, Charles and Petropoulos, P. Marios and Siampos, Konstantinos",
    title = "{Two-dimensional fluids and their holographic duals}",
    eprint = "1812.04019",
    archivePrefix = "arXiv",
    primaryClass = "hep-th",
    reportNumber = "CPHT-RR078.082018, CERN-TH-2018-231",
    doi = "10.1016/j.nuclphysb.2019.114692",
    journal = "Nucl. Phys. B",
    volume = "946",
    pages = "114692",
    year = "2019"
}

@article{Donnay:2022wvx,
    author = "Donnay, Laura and Fiorucci, Adrien and Herfray, Yannick and Ruzziconi, Romain",
    title = "{Bridging Carrollian and celestial holography}",
    eprint = "2212.12553",
    archivePrefix = "arXiv",
    primaryClass = "hep-th",
    doi = "10.1103/PhysRevD.107.126027",
    journal = "Phys. Rev. D",
    volume = "107",
    number = "12",
    pages = "126027",
    year = "2023"
}

@article{Bianchi:2001kw,
    author = "Bianchi, Massimo and Freedman, Daniel Z. and Skenderis, Kostas",
    title = "{Holographic renormalization}",
    eprint = "hep-th/0112119",
    archivePrefix = "arXiv",
    reportNumber = "MIT-CTP-3166, PUTP-1999, DAMTP-2001-63, ROM2F-2001-30",
    doi = "10.1016/S0550-3213(02)00179-7",
    journal = "Nucl. Phys. B",
    volume = "631",
    pages = "159--194",
    year = "2002"
}

@article{Skenderis:2002wp,
    author = "Skenderis, Kostas",
    editor = "de Wit, B. and Vandoren, S.",
    title = "{Lecture notes on holographic renormalization}",
    eprint = "hep-th/0209067",
    archivePrefix = "arXiv",
    reportNumber = "PUTP-2047",
    doi = "10.1088/0264-9381/19/22/306",
    journal = "Class. Quant. Grav.",
    volume = "19",
    pages = "5849--5876",
    year = "2002"
}

@article{Mason:2023mti,
    author = "Mason, Lionel and Ruzziconi, Romain and Yelleshpur Srikant, Akshay",
    title = "{Carrollian amplitudes and celestial symmetries}",
    eprint = "2312.10138",
    archivePrefix = "arXiv",
    primaryClass = "hep-th",
    doi = "10.1007/JHEP05(2024)012",
    journal = "JHEP",
    volume = "05",
    pages = "012",
    year = "2024"
}

@article{Alday:2024yyj,
    author = "Alday, Luis F. and Nocchi, Maria and Ruzziconi, Romain and Yelleshpur Srikant, Akshay",
    title = "{Carrollian Amplitudes from Holographic Correlators}",
    eprint = "2406.19343",
    archivePrefix = "arXiv",
    primaryClass = "hep-th",
    month = "6",
    year = "2024"
}

@article{Donnay:2023mrd,
    author = "Donnay, Laura",
    title = "{Celestial holography: An asymptotic symmetry perspective}",
    eprint = "2310.12922",
    archivePrefix = "arXiv",
    primaryClass = "hep-th",
    doi = "10.1016/j.physrep.2024.04.003",
    journal = "Phys. Rept.",
    volume = "1073",
    pages = "1--41",
    year = "2024"
}

@article{deBoer:1999tgo,
    author = "de Boer, Jan and Verlinde, Erik P. and Verlinde, Herman L.",
    title = "{On the holographic renormalization group}",
    eprint = "hep-th/9912012",
    archivePrefix = "arXiv",
    reportNumber = "PUPT-1898, ITFA-99-39, SPIN-1999-29",
    doi = "10.1088/1126-6708/2000/08/003",
    journal = "JHEP",
    volume = "08",
    pages = "003",
    year = "2000"
}

@article{Compere:2020lrt,
    author = "Comp\`ere, Geoffrey and Fiorucci, Adrien and Ruzziconi, Romain",
    title = "{The $\Lambda$-BMS$_4$ charge algebra}",
    eprint = "2004.10769",
    archivePrefix = "arXiv",
    primaryClass = "hep-th",
    doi = "10.1007/JHEP10(2020)205",
    journal = "JHEP",
    volume = "10",
    pages = "205",
    year = "2020"
}

@article{Gadioux:2023pmw,
    author = "Gadioux, Maxime and Reall, Harvey S.",
    title = "{Creases, corners and caustics: properties of non-smooth structures on black hole horizons}",
    eprint = "2303.15512",
    archivePrefix = "arXiv",
    primaryClass = "gr-qc",
    month = "3",
    year = "2023"
}

\end{document}